# Multifunctional Magnetoelectric Materials for Device Applications

N. Ortega[1], Ashok Kumar[1], J.F. Scott[1,2] and Ram S. Katiyar[1]

[1]Department of Physics and Institute for Functional Nanomaterials, University of Puerto Rico, San Juan, PR 00931-3343 USA
[2]Department of Physics, University of Cambridge, Cambridge CB2 3EQ, UK

## Abstract

Mutiferroics are a novel class of next generation multifunctional materials, which display simultaneous magnetic spin, electric dipole, and ferroelastic ordering, and have drawn increasing interest due to their multi-functionality for a variety of device applications. Since single-phase materials exist rarely in nature with such cross-coupling properties, an intensive research activity is being pursued towards the discovery of new single-phase multiferroic materials and the design of new engineered materials with strong magneto-electric (ME) coupling. This review article summarizes the development of different kinds of multiferroic material: single-phase and composite ceramic, laminated composite, and nanostructured thin films. Thin-film nanostructures have higher magnitude direct ME coupling values and clear evidence of indirect ME coupling compared with bulk materials. Promising ME coupling coefficients have been reported in laminated composite materials in which signal to noise ratio is good for device fabrication. We describe the possible applications of these materials.

## Introduction

Ferroelectric (FE) and ferromagnetic (FM) materials have been widely studied and have led to important scientific and technological advances. Different origins of ferroelectricity have been explored, such as charge ordering, magnetically induced ferroelectricity, lone-pair electron effects, octahedral distortion, strain mediation, geometrical frustration, etc., but in general, ferroelectricity results from relative shifts of negative and positive ions that induce surface charges; usually it requires formally empty $d$ orbitals, but on the other hand, magnetism is related to ordering of spins of electrons in incomplete ionic shells, that is, results from partially filled $d$ orbitals. Ferroelectric materials are used to make capacitors with high dielectric constant, transducers and actuators, which are not switching devices; and their memory applications exploit their hysteresis properties which result in two stable state of opposite polarization. Similarly, ferromagnetic materials are used in many industrial applications, such as sensors, read heads, and memories based on giant magnetoresistive effects, which they are already commercially available [1]. Another novel class of multifunctional materials, known as **magnetoelectric (ME) multiferroics (MF)** possess simultaneously **ferroelectric** and **ferromagnetic** properties in the same phase and exhibit linear coupling between them, have had a strong research interest due to both the basic physics and the number of potential multi-functional applications in modern technologies. Such materials display the magnetoelectric (ME) coupling in which the spontaneous magnetization can be switched by an applied electric field and the spontaneous electrical polarization that can be switched by an applied magnetic field, due to the cross coupling between FE and FM ordering. Multiferroic are likely to offer additional functional parameters including more than two logic states for a whole range of new applications [2]. These ME materials are often combinations that have all the potential properties of their parent FE and FM materials; in addition they have bifunctional properties not available in systems with a single order parameter. The search for these materials is driven by the prospect of controlling charges by applied magnetic fields and spins by applied voltages and using this to construct new forms of multifunctional devices [3,4]. The coexistence of magnetization and electric polarization might allow an additional degree of freedom in the design of novel devices such as actuators, transducers and storage devices. Other potential applications include multiple-state memory elements, in which data is stored both in the electric and the magnetic polarizations, or novel memory media which might allow writing of the ferroelectric data bit, and reading of the magnetic field generated by association [5].

Single-phase multiferroic materials have been reported in small numbers due to the contra-indication mentioned above between the conventional mechanism in ferroelectric oxides that requires empty $d$-orbitals and formation of magnetic moments which results from partially filled $d$-orbitals [1,6,7]. Alternatively the fabrication of artificial multiferroics includes the possibility of tailoring the properties using two different compounds, one being ferromagnetic and other being ferroelectric. The aim of this technique is to create materials that display properties of the parent compounds and their coupling. These composite multiferroic ceramics have been found to exhibit a larger ME effect than that of the single-phase materials by more than one order of magnitude [8]. However, the advances in thin-film growth techniques have provided routes to structures and phases that are inaccessible by chemical methods, generating attention to composite ME materials, especially thin films. Nano-structuring is a promising approach, which has opened the door to the design of practical devices based on ME coupling. Thin-film multiferroics begin to reveal a range of fascinating phenomena as well as to stimulate exploration the new device heterostructures, which have potential applications in microdevices and integrated units such as microsensors, micro-electro-mechanical systems (MEMS) devices, and high-density information storage. Several efforts have been made to synthesize ME composite thin films where FE and FM properties coexist for the next generation of miniaturized integrated devices. Different configurations were reported for fabrication of ME composite thin films such as: composites with FE or FM terminal layers, double multilayers, superlattices consisting of alternating FM and FE layers, epitaxial ferroelectromagnetic nanocomposites by self-assembly technique, etc. These studies clearly demonstrated the ME effect in different configurations [9].

Each stage of development of multiferroic materials will be presented in more detailed subsequent sections of this article along with some examples of newly developed materials in each class of multiferroic that has been investigated and recently published. The last part of the review article will be devoted to the possible applications of magnetoelectric multiferroic materials.

**Background and history of multiferroic and magnetoelectric**

**Bulk Single Phase Multiferroic:** Two independent events mark the birth of the ME: i) In 1888 Röntgen discovered that a moving dielectric became magnetized when placed in an electric field [10]; the reverse effect was observed 17 years later (polarization of a moving dielectric in a magnetic field) [11]. ii) In 1894 Curie pointed out the possibility of intrinsic ME behavior of crystals on the basis of symmetry considerations, in which the crystals can be polarized in the presence of a magnetic field and vice-versa. Later, Dzyaloshinskii provided theoretical details of the ME in a specific material -- $Cr_2O_3$[12]; subsequently in 1960 Astrov reported the first experimental observation of ME effect in $Cr_2O_3$, and he showed experimental confirmation of an electric field induced magnetization [13,14], and Rado et al. [15,16] quickly confirmed the converse magnetic field induced polarization. The subsequent search for alternative ME materials led scientists to synthesize new single phase compounds and found multiferroic and ME behavior in several families of materials such as: Pb-family[17,18,19,20,21,22,23,24] perovskite oxide structure; Bi compounds [1,25,26,27,28]; rare earth (RE) manganites [29,30,31,32]; mixed perovskite solid solutions [33,34,35,36,37,38,39,40]; $REMn_2O_5$ family [41,42,43,44,45,46,47], phosphates[48,49], boracites[50,51,52,53,54], fluoride family[55,56,57,58,59,60], spinel chalcogenides [61,62,63,64,65,66], and delafossites, [67,68,69]. Table 1 summarizes the characteristics of each family, some typical examples, and references. Most of the compounds listed in Table 1 have drawbacks: i) they are multiferroic at low temperature; ii) they have very low ferroelectric and/or ferromagnetic response; and iii) low value of magnetoelectric coupling constant for practical applications. Several review articles summarize the general aspects of single phase multiferroic materials and the ME coupling effect [7,70,71,72].

Table 1. Single phase multiferroic materials

| Material | Characteristic | Examples | Ref. |
|---|---|---|---|
| | **Single phase Multiferroic** | | |
| **Perovskite oxides** | **Relaxor**: FE, anti-FM and weak ferromagnetism. | $Pb(Fe_{0.5}Nb_{0.5})O_3$ (PFN) $Pb(Fe_{0.5}Ta_{0.5})O_3$ (PFT) $Pb(Fe_{2/3}W_{1/3})O_3$ (PFW) | 17-24 |
| | **Bismuth compounds**: weakly anti-FM, ($Fe^{3+}$ with its $3d^5$ electrons provide the magnetism) and FM (polarization of $6^2$ lone pair of $Bi^{3+}$) | $BiMnO_3$ $BiFeO_3$ | 1,25-28 |
| | **The perovskite rare earth (RE) manganites** ($REMnO_3$) RE = (Y, Ho, Er, Tm, Yb, Lu, Sc, Tb) Anti-FM, and FE. The ME coupling results from weak spin-orbit interactions. | $ErMnO_3$ $Yb\,MnO_3$ $Y\,MnO_3$ $Tb\,MnO_3$ $Lu\,MnO_3$ | 29-32 |
| | **Mixed Perovskite solid solution:** Partially replace diamagnetic ions by paramagnetic ions on the B-site of oxyoctahedral perovskite cause the FE (Polarization of $6s^2$ lone pair of Pb) and the formally $d^5$ $Fe^{3+}$ ion is responsible for the magnetic ordering. | $PFW$-$PbTiO_3$ $PFW$-$Pb(Mg_{1/2}W_{1/3})O_3$ $PFN$-$Pb(Zr_{0.2}Ti_{0.8})O_3$ $PFN$-$Pb(Zn_{1/3}Nb_{2/3})O_3$ $PFT$-$Pb(Zr_{0.53}Ti_{0.47})O_3$ | 33-40 |
| **Other oxides** | **$REMn_2O_5$** (RE = Y, Tb, Dy, Ho…) The ME coupling is mediated by strong super exchange not weak spin orbit interactions. The ME leads that the electric field control of spin chirality. | $TbMn_2O_5$, $DyMn_2O_5$ $HoMn_2O_5$, $GdMn_2O_5$ $EuMn_2O_5$, $YMn_2O_5$ | 41-47 |
| **Non oxides** | **Phosphates:** $LiMPO_4$ (M = Ni, Co, Fe, Mn). Weak FM, magnetoelectricity, magnetic incommensurate and high magnetic field induced phase transition. | $LiNiPO_4$ $LiFePO_4$ $LiCoPO_4$ $LiMnPO_4$ | 48-49 |
| | **Boracites** $M_3B_7O_{13}X$, where M = (Ni, Cu, Cr, Mn, Fe, Co) and X = (Cl, Br, I), FE, ferroelastic, anti-FM and weak FM (at low temperature) properties | $Ni_3B_7O_{13}I$ $Co_3B_7O_{13}Cl$ | 50-54 |
| | **Fluoride: $BaMF_4$** M = Mg, Mn, Fe, Co, Ni, Zn. Spontaneous ME coupling permits the FE order to cant the spins of the two anti-aligned magnetic sublattices along the axis of the crystal, giving a small net FM magnetization. | $BaMnF_4$ $BaNiF_4$ $BaFeF_4$ | 52-57 |
| | **Spinel Chalcogenides** Low temperature multiferroic. Colossal magnetocapacitance due to in large part for extrinsic effects (magnetoresistance). | $ZnCr_2Se_4$ $CdCr_2S_4$ | 61-66 |
| | **Delaffosite structure** Magnetic-field-induced collinear-noncollinear magnetic phase transitions. Magnetic, ME, and magnetoelastic measurements reveal that noncollinear helimagnetic structure plays an essential role in inducing electric polarization. | $CuFeO_2$ $CuCrO_2$ | 67-69 |

FE: Ferroelectric, FM: Ferromagnetic, ME: Magnetoelectric

# Ceramic Single Phase Multiferroics

## A New Room-temperature Multiferroic $(PbZr_{0.53}Ti_{0.47}O_3)_{(1-x)}$- $(PbFe_{0.5}Ta_{0.5}O_3)_x$

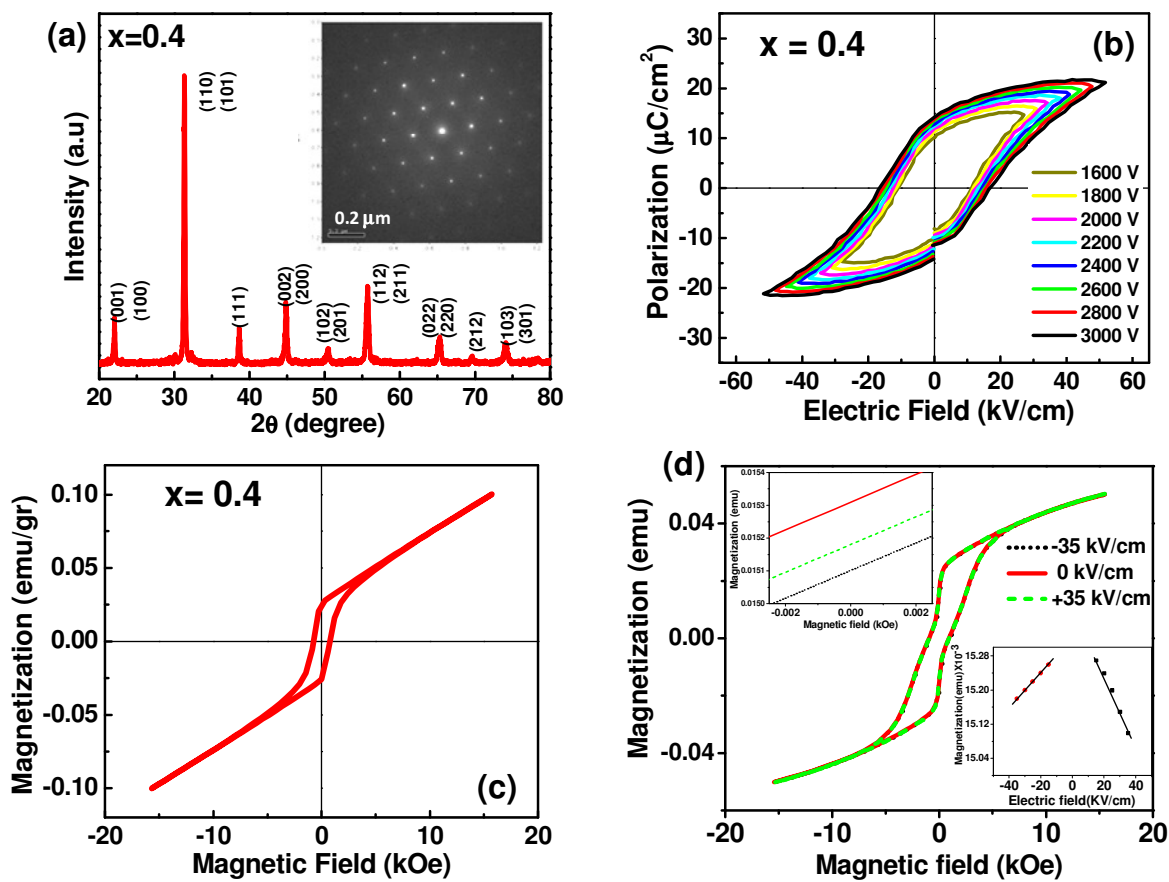

Figure 1. PZTFT x= 0.4 multiferroic material (a) X-ray diffraction patterns and selected area electron diffraction (SAED) patterns show single phase material (b) ferroelectric and (c) ferromagnetic hysteresis loop sowing multiferroic behavior. (d) Electric control of magnetization, their linear ME coupling (lower inset), and change in remanent magnetization under external field (upper inset). Adapted from Ref. [39], reproduced with permission (© 2011 AIP).

Perovskite solid solutions have been an active area of research due to possibility of exploring new materials with multifunctional applications [18,19]; Sanchez et al. [39] combined the best qualities of both lead iron tantalate with those of PZT synthesizing $(PbZr_{0.53}Ti_{0.47} O_3)_{(1-x)}$- $(PbFe_{0.5}Ta_{0.5}O_3)_x$ (PZTFT). PZTFT with variable x was prepared by conventional solid-state route. PZTFT is a single-phase (Figure 1(a)), low-loss material, it showed both ferroelectric (Figure 2(b)) and ferromagnetic (Figure 2(c)) ordering at room temperature with magnetoelectric coupling (Figure 2(d)), with very low leakage current even at 550 K, such that perfect square hysteresis at 550 K was observed. Temperature-dependent XRD revealed a structural phase transition from tetragonal to orthorhombic under cooling. PZTFT shows classic textbook example of several sequential phase transitions: cubic-tetragonal (ca. 1300 K) to tetragonal- orthorhombic (520 K for x= 0.3 and 475 K for x= 0.4); and to orthorhombic-rhombohedral (230 - 270 K) – a sequence similar to that in barium titanate. ME coupling is of the same order of magnitude as in the single phase multiferroic PFT. High resolution transmission electron microscope (HRTEM) studies are being carried out to verify the possibility of secondary phase or cluster at nanoscale level. Inset from Figure 2(a) show the of the selected area electron diffraction (SAED) pattern of the ceramic PZTFT, these studies suggest no evidence of secondary diffraction pattern in the PZTFT material at nanoscale. PZTFT multiferroic is the low-loss temperature multiferroic known, which is a great advantage for magnetoelectric devices.

**Bulk Composite Multiferroics**

In order to compensate the deficiencies of the natural multiferroics at ambient temperatures the fabrication of artificial multiferroics for enhance magnetoelectric effects opens the possibility of tailoring the properties using two different compounds, one being ferromagnetic and other being ferroelectric [8,26]. The aim of this technique is to create materials that display both properties of the parent compounds and their cross coupling, but the cross coupling is generally indirect via strain (magnetostriction plus electrostriction) and not direct P.M interaction. A magnetic field applied to the composites will induce magnetostriction which produces piezoelectric and/or electrostrictive response. This ME property of the ferroelectric-ferromagnetic composite is known as a product-property of the composite [73]. The ME composites could have various connectivity schemes, but the common connectivity schemes are: i) **0-3** type particulate composite where the piezoelectric/magnetic particles are embedded in a matrix of magnetic/piezoelectric phases; ii) **2-2** type laminate ceramic composites consisting of the of piezoelectric and magnetic oxide layers; and iii) **1-3** type "fibers" of one phase embedded in the matrix of another phase [8]. This composite approach opens new avenues for tailoring the ME response through the choice, phase properties, volume fraction, shape, connectivity and microstructure of the constituents; indeed room temperature ME coupling coefficients have been achieved that exceed the low temperature values found in single phase compounds by three to five orders of magnitude [70].
This idea was initially developed in ceramic-composite (0-3 type) [74-93] and later laminated multilayers have been investigated [94-104].

The first work on the ME composite was done at the Philips laboratory [74, 75-78] the ME composites were prepared by unidirectional solidification of a eutectic composition of the quaternary system Fe-Co-Ti-Ba-O, the ME coefficient $dE/dH = 130$ mV cm$^{-1}$ Oe$^{-1}$[74,76], which was superior to the single phase materials such as $Cr_2O_3$. In 1978, they reported on the sintered ME composite of $BaTiO_3$ and $Ni(Co,Mn)Fe_2O_4$ with excess $TiO_2$ in terms of the particle size effect, the cooling rate, and the mole ratios of both the phases [77]. After the pioneering experiment on BTO/CFO [74-77], a variety of ferroelectric/ferrite bulk compositions have been fabricated consisting of the ferroelectric phases, such as $BaTiO_3$ (BTO), $PbZr_{1-x}Ti_xO_3$ (PZT), , $PbTiO_3$ (PT), polyvinylidenefluoride (PVDF) and ferromagnetic phases such as $CoFe_2O_4$ (CFO), $Tb_{1-x}Dy_xFe_2$-Terfenol-D, $NiFe_2O_4$ (NFO), $CuFe_2O_4$ (CuFO), $LaMnO_3$ (LMO). Although bulk composite materials were considered to exhibit promising (larger) ME effects, so far what is observed in ceramic composites is ten times lower than that predicted, due mainly to their inherent preparation problems, such as atomic interfacial diffusion, chemical reactions between the constituent starting materials during sintering processes or interdiffusion of the phases. However, important efforts have been made to obtain the desired properties of individual constituents such as ferroelectric (good piezoelectric, high electromechanical coupling coefficients) and ferrite phases (good piezomagnetic, high magnetomechanical coupling coefficients, high resistivity), as well as the improvement of parameters related to the composite materials, such as carefully chosen constituent materials, sintering processes, grain size, mole ratios, thermal expansion mismatch between phases, and uniform distribution of magnetostrictive phase within a piezoelectric matrix.

Table 2. Composite multiferroic materials

| **Composite Multiferroic** ||||| 
| Material | Characteristic | Examples | $\alpha$ (mV cm$^{-1}$ Oe$^{-1}$) | Ref. |
|---|---|---|---|---|
| **Particulate Ceramic Composite** | A variety of ferroelectric/ferrite compositions have been fabricated consisting of the ferroelectric phases, such as BaTiO$_3$ (BTO), PbZr$_{1-x}$Ti$_x$O$_3$ (PZT), polyvinylidenefluoride (PVDF), PbTiO$_3$ (PT) and ferromagnetic phases such as CoFe$_2$O$_4$ (CFO), Terfenol-D (Tb$_{1-x}$Dy$_x$Fe$_2$), NiFe$_2$O$_4$ (NFO), CuFe$_2$O$_4$ (CuFO), LaMnO$_3$ (LMO), | BTO/CFO<br>CuFO/ BTO<br>CFO/ BTO<br>PZT/CFO<br>Pb$_{0.95}$Sr$_{0.05}$Ni$_{0.06}$Zr$_{0.49}$Ti$_{0.40}$O$_3$/CFO<br>NFO/PZT<br>BTO/LMO<br>Terfenol-D/ PVDF<br>BTO/Ni$_{0.94}$Co$_{0.01}$Cu$_{0.05}$Fe$_2$O$_4$<br>Sr$_{0.5}$Ba$_{0.5}$Nb$_2$O$_6$/CFO<br>ErMnO$_3$/La$_{0.7}$Sr$_{0.3}$MnO$_3$<br>CoFe$_2$O$_4$–BiFeO$_3$<br>Ni$_{0.83}$Co$_{0.15}$Cu$_{0.02}$Fe$_{1.9}$O$_4$/PZT<br>0.9Pb(Zr$_{0.52}$Ti$_{0.48}$)O$_3$-0.1Pb(Zn$_{1/3}$Nb$_{2/3}$)O$_3$-Ni$_{0.8}$Zn$_{0.2}$Fe$_2$O$_4$ | 130<br>425<br>200<br>----<br>70<br>80-115<br>-----<br>1430<br>0.64<br>25<br>----<br>285<br>3150<br>75 | 74,75,76,77<br>78,79<br>80<br>81<br>82<br>83-84<br>85<br>86<br>87<br>88<br>89<br>90<br>91<br>92 |
| **Laminated Composite** | The mechanical stress mediating between the electrical and magnetic properties of the composite should be passed through the constituents with losses as low as possible, These obstacles were overcome in 2001 by using laminar instead of particulate composites. | PZT/NiFe$_2$O$_4$ (bilayer)<br>PZT/NiFe$_2$O$_4$ (multilayer)<br>PZT-CFO<br>PZT/La$_{0.7}$Sr$_{0.3}$MnO$_3$<br>PZT/La$_{0.7}$Ca$_{0.3}$MnO$_3$<br>Terfenol-D (T)-PVDF/PZT(P)-PVDF/T T/P/T<br>PZT rod arrays embedded in Terfenol-D<br>PZT/ Terfenol-D discs<br>PVDF-hexafluoropropylene-Metglass | 20, 460$^T$-45$^L$<br>1500<br>55$^T$-5$^L$,<br>160$^T$-30$^L$<br>60$^T$,20$^L$<br>3000-4000<br>6000<br>6000<br>4680<br>12000 | 93,94,95<br>95<br>95,96<br>97<br>97<br>98,99<br>100<br>101<br>102<br>103 |

For example, Ryu et al. [92] recently published improvement of ME properties with more homogeneous dispersion of a magnetostrictive phase (Ni$_{0.8}$Zn$_{0.2}$Fe$_2$O$_4$) into a piezoelectric matrix 0.9Pb(Zr$_{0.52}$Ti$_{0.48}$)O$_3$-0.1Pb(Zn$_{1/3}$Nb$_{2/3}$)O$_3$ (see Figure 2(a) and 2(b)) and Ramanaa et al. [91] synthesized Ni$_{0.83}$Co$_{0.15}$Cu$_{0.02}$Fe$_{1.9}$O$_4$ (NCCF)/PZT; they add Co and Cu to NiFeO$_3$ to increase the resistivity of the ferrite phase (NCCF), obtaining a maximum ME coefficient of 3.15 Vcm$^{-1}$Oe$^{-1}$ in the composite containing 0.5NCCF + 0.5PZT, as is shown in Figure 2(c). Due to the enriched engineering process and technologically, a viable ME response can be observed in multiferroic ME composites above room temperature. Various ME bulk composites in different systems have been investigated; the first part of Table 2 show the list of some bulk composite, that have been investigated over the last forty years.

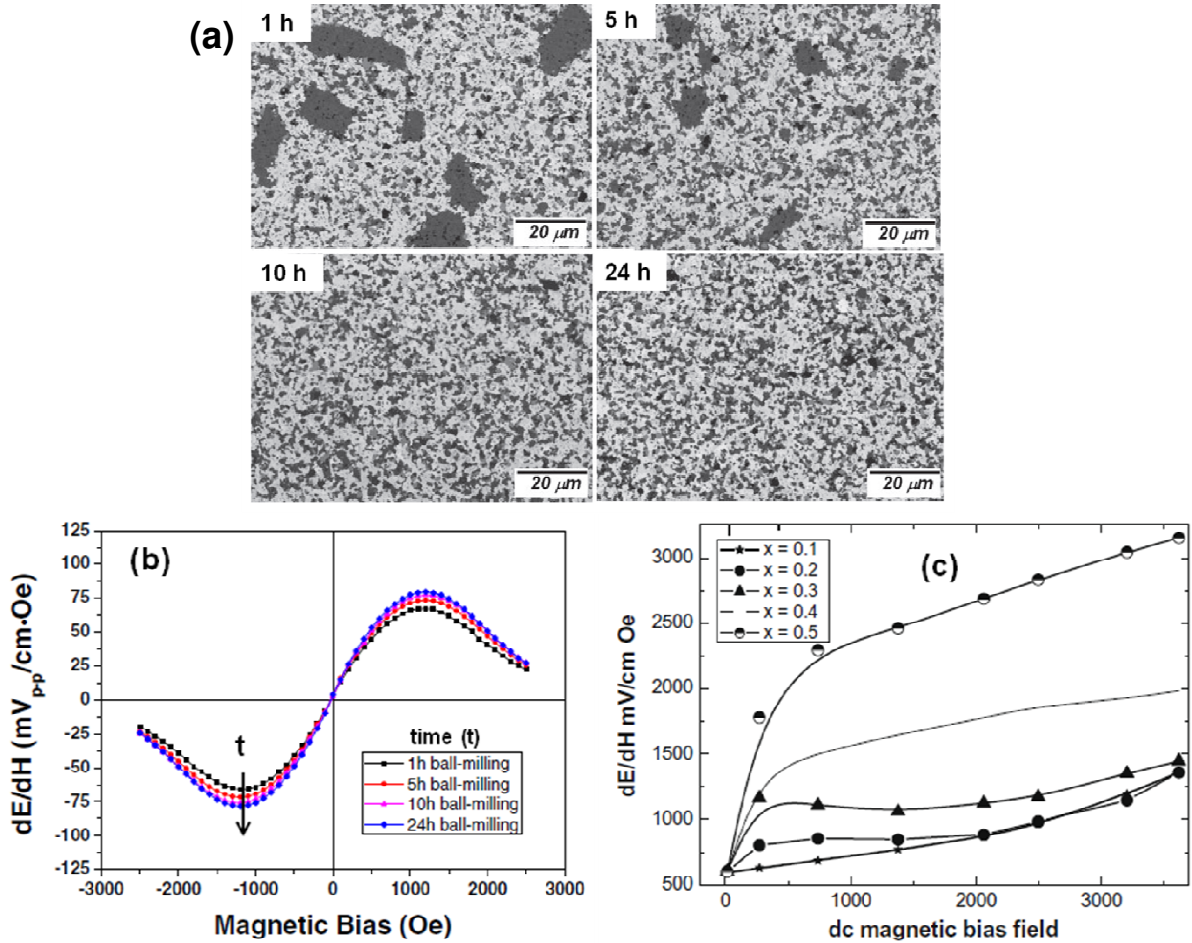

Figure 2. (a) Backscattering emission image obtained by SEM for the PZT–PZN-NZF composite sintered at 1200 °C for 2 h with different NZF ball-milling times. (b) (Color online) Magnetoelectric properties (dE/dH) of PZT–PZN-NZF composite sintered at 1200 °C for 2 h. Adapted from Ref. [92] reproduced with permission (© 2011 The Japan Society of Applied Physics). (c) ME voltage coefficient as a function of applied dc magnetic bias field for the composites (1-x)PZT – xNCCF. Adapted from Ref. [91] reproduced with permission (© 2009 Elsevier).

**Laminated Composite Multiferroic**

However, in spite of all these efforts considerable enhancement of the composite ME response beyond ~3000 mV cm$^{-1}$ Oe$^{-1}$ was not attained in almost four decades of research, although the observed ME coefficients was less than 1–2 orders of magnitude compared with the predicted values by theoretical works. Several reasons for this discrepancy were already mentioned previously. ME properties of the composite materials was not good enough to be used in practical applications. Using laminar composite materials instead of particulate composites some problems were overcome [70].
.
Laminated composite have much higher ME coefficient than that of single phase materials or particulate composites. Another remarkable difference between the laminated and particulate composite ceramics is that the laminated structures exhibit much larger anisotropy then the particulate one [8]. ME couplings have been studied experimentally [74-103] and theoretically [104,105,106]. Potentially these new materials and structures can be used for magnetic field sensors, current sensors, energy harvesters, transformers; ME filters, as well as phase shifters. In laminated composites the ME response strongly depends on the interfacial bonding between two layer, since the mechanical stress mediating between the dielectric and magnetic properties of the composite should be passed through the constituents with losses as low as possible. Throughout the years various efforts have been made to improve the ME coefficient values in the laminar composite, these efforts have been directed to preparation techniques of the sample, the choice of materials, different structures and thickness of the sample. Initially the technique used to fabricate the laminated composite was conventional sintering; however, in order to suppress inter-diffusion and chemical reaction of the constituent and improve the quality of the laminate composite, the hot pressing technique has been employed, i.e. Nan's group reported the increase of the ME coefficient up to 6000 mVOe$^{-1}$cm$^{-1}$ when the Terfenol-D-PVDF/PZT-PVDF/ Terfenol-D-PVDF (T/P/T) layer structures were fabricated using a hot molding press (see Figure 3) [98,100].

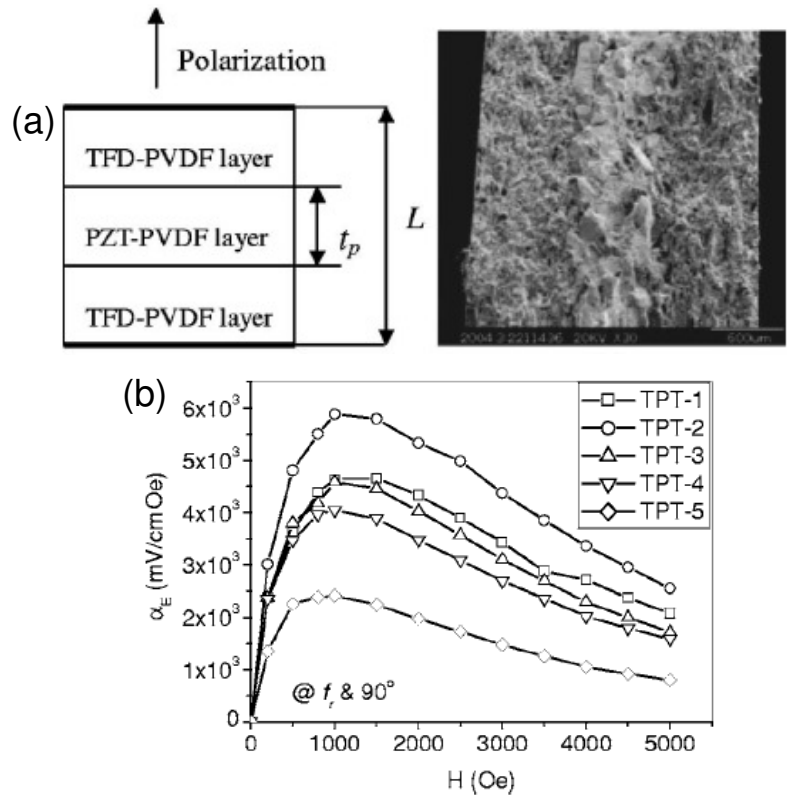

Figure 3. (a) Schematic illustration and the typical micrograph of the fractured surface of the laminated *TPT* composites. TFD denotes Terfenol-*D*. (b). The ME sensitivity $\alpha_E$ values measured at resonance frequency ($f_r$) for the laminated composites as a function of the bias magnetic field at the measuring angle $\theta = 90°$. In the legend 1,2,3,4,5 correspond to 1/7, 2/7, 3/7, 4/7, 5/7 ($t_p/L$) ratio respectively, for various TPT laminated samples. Adapted from Ref. [100] reproduced with permission (© 2005 APS).

Different type of materials can be used in fabrication of ME laminate composite, i.e hard materials as piezoelectric (PZT, BTO) and piezomagnetic (NFO,CFO, LSMO, LSCO) ceramics; another class is soft materials, in case of the piezomagnetic we can mention Metglass and Terfenol-D ( $Tb_xDy_{1-x}Fe_2$ (x ~ 0.3)); and as soft piezoelectric polymers we have polyvinylidenefluoride (PVDF), P(TrFE). Large ME responses have been obtained with the combination of hard and soft materials in the laminated composite.

The most common studied laminated composite structure is the 2-2 structure as bilayers or multilayers [93-97]; however, quasi 2-2 structures can be obtained in case of Terfenol-D-polymer (PVDF) and PZT–polymer (PVDF): in this case the PVDF polymer is used just as a matrix binder [98-100]. Another important structure is 1-3; Shi et al.[101] prepared this via a dice- and fill process where the PZT bulk is diced to get a PZT rod array, and then the gap of the PZT rod array is filled with a mixture of Terfenol-D particles and epoxy. When the epoxy hardens, the pseudo 1-3 type multiferroic composite is obtained. Recently large ME coupling (12000 $mVOe^{-1}Cm^{-1}$) was achieved in PVDF-hexafluoropropylene-Metglass fabricated by extrusion-blown and a hot-press quenched method; an additionally interesting phenomenon -- a field-induced phase transition -- was observed in this system (see Figure 4). A list of several laminated composites appear in Table 2 with their respective ME coefficient values shows the development in this area and emphasizes the superior ME coefficient values obtained in laminar composite compared with ceramic composite and single crystal.

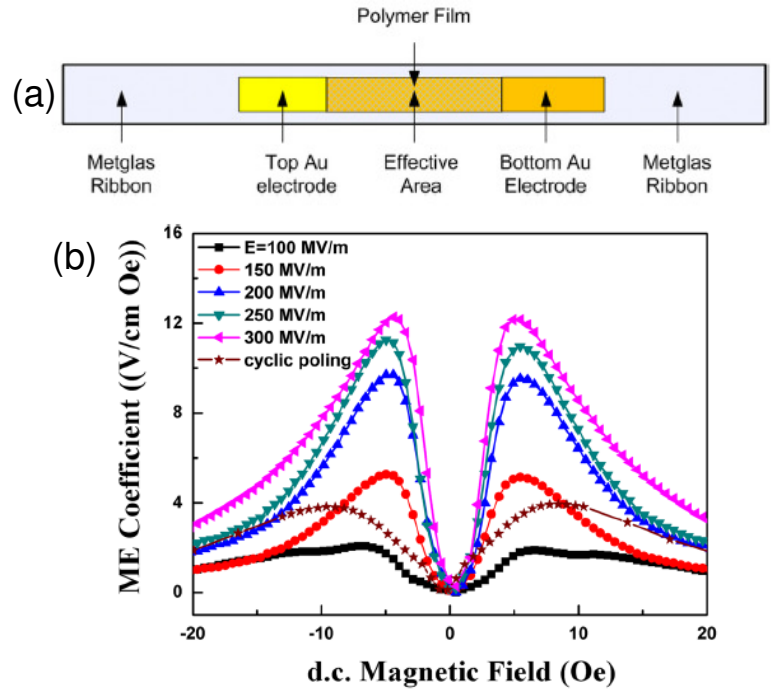

Figure 4. (a) Schematic configuration of P(VDF-HFP)/Metglas laminate. The polymer film was smaller in area than the Metglas and placed at the central area, optimizing the magnetic flux effect. (b) ME coefficient of hot-pressed P(VDF-HFP)/Metglas laminates as a function of d.c. magnetic field for conventional poling. The ME coefficient vs d.c. magnetic field for cyclic poling ($E_p$ = 225 V/m for extruded film) is also shown. Adapted from Ref [103]; reproduced with permission (© 2011 AIP).

**Single and Heterostructure Thin Films**

The advances in thin-film growth techniques provided a controlled way of synthesizing high quality single and nano-structures thin films and have allowed the properties of existing materials to be modified by strain engineering [25,107], on the other hand nano-structuring is a promising approach, which has opened the door to the design of practical devices based on ME coupling. Multiferroic single phase and nanostructure thin films have been produced using a wide variety of grown techniques including sputtering, spin coating, metal-organic chemical vapor deposition (MOCVD), sol-gel process, pulsed laser deposition (PLD), molecular beam epitaxy (MBE), etc. Research interest in magnetoelectric multiferroic thin films has been increasing the last decade, due to the new range of phenomena observed, which have potential applications in micro-devices and integrated units such as micro-sensors, micro-electro mechanical systems (MEMS) devices, and high-density information [2]. In case of single-phase thin film multiferroics most of the published work is devoted to hexagonal manganites such as $YMnO_3$ (YMO) [108-119] and Bi-based perovskite (mostly pure or chemically doped $BiFeO_3$) [120-133]; another series of single phase materials are solid solution of lead based solid solution perovskite [37,134-136] ME materials has also been investigated.

**Single-Phase Multiferroic Thin Films**

**Yttrium manganite, $YMnO_3$ (YMO):** This $ABO_3$ oxide belongs to the $RMnO_3$ (R = rare-earth element) system which can be divided into two subsystems, in which the crystal structure is mainly determined by the size of R cation. For the larger R size (R = La, Dy) an orthorhombic perovskite structure is adopted, whereas for the smaller R (R= Sc, Y,Ho, Lu) the structure is hexagonal. One of the typical rare-earth manganite studied in thin-film form is $YMnO_3$, its structure can be hexagonal (ferroelectric) or orthorhombic (non-ferroelectric) depending on epitaxial strain. Hexagonal YMO undergoes a paraelectric-ferroelectric phase transition at $T_c \sim$ 900 K, and it display an antiferromagnetic transition around 70 K; that is, YMO is a room-temperature ferroelectric and low-temperature multiferroic [108]. Aken et al.[109] reported that the ferroelectricity mechanism in YMO is driven entirely by electrostatic and size effects (dipole-dipole interaction and oxygen rotations), rather the usual changes in chemical bonding associated with FE phase transition in perovskite oxide; although a complete analysis of the phase transitions sequence of YMO with the possible existence of intermediate phases i.e *$P6_3mc$* (FE) and *$P6_3/mcm$* (paraelectric) between *$P6_3cm$* (FE) and *$P6_3/mmc$* (paraelectric) was reported subsequently by Fennie and Rabe [110] . The FE thin films are expected to be more suitable dielectric materials for metal-ferroelectric-semiconductor field-effect transistor (MFSFET) structures due to absence of volatile components as Pb, Ti or Ta, which can easily diffuse into semiconductors. Fujimura et al. [111], proposed $YMnO_3$ thin films as a new candidate for nonvolatile memory devices, using rf magnetron sputtering, they fabricated epitaxial (0001) $YMnO_3$ films on (111) MgO, (0001)ZnO:Al/(0001) sapphire, and polycrystalline films on (111)Pt/(111)MgO. The dielectric properties of the epitaxial and polycrystalline $YMnO_3$ films were almost the same. After Fujimura's publication in 1996, YMO thin films have been synthesized via a wide variety of substrates and growth techniques [107]. Lee et al. [112] used rf sputtering to grow c-axis oriented Pt/YMO/$Y_2O_3$/Si and Pt/YMO/Si -- that is, metal/ferroelectric/insulator/semiconductor (MEFIS) and metal/ferroelectric/ semiconductor (MFS) structures, respectively. Electrical properties showed the memory window of the MEFIS is greater than that of the MFS capacitor. Later, Yoo et al., [113] using the same growth technique and substrate utilized in reference 112, found the c-axis/polycrystalline (bi-layer)

YMO thin films showed a better memory window than both purely c-axis oriented and purely polycrystalline YMO thin films. Epitaxial (0001) YMO thin films, on (111)Pt/(0001)sapphire (epi-YMO/Pt) and $Y_2O_3$/Si (epi/YMO/Si) and YMO oriented polycrystalline films on (111)Pt/$ZrO_2$/$SiO_2$/Si were grown by PLD technique by Ito et al.[114]; excellent ferroelectric properties was obtained in epitaxial films compared to oriented polycrystalline films. The crystallinity and CV hysteresis of the epi-YMO/Pt and epi/YMO/Si thin films were almost the same. Conclusion: Epitaxial YMO thin film is a ferroelectric material suitable for MEFIS applications. Using also PLD, Dho [115] et al. found the growth orientation and surface morphology of YMO films were sensitively dependent on the substrate (lattice mismatch) and deposition conditions (oxygen background pressure); that is, for (111) $SrTiO_3$ (STO) substrates, the orthorhombic YMO phase competing with the hexagonal phase, which was induced by the strong compressive stress. YMO thin films were also synthesized by sol-gel process using alkoxides on Pt(111)/Ti/$SiO_2$/Si [116]. It was found that the higher drying temperature affects the ferroelectric properties due to higher crystallinity with the c-axis preferred orientation; as a result, the highly c-axis oriented $YMnO_3$ thin films exhibited higher remanent polarization ($2P_r$=3.6 μ/$cm^2$) compared with randomly oriented $YMnO_3$ thin films. More recently, Uusi-Esko et al. [117] used atomic layer deposition (ALD) to grown both the hexagonal and orthorhombic forms of $YMnO_3$. On Si(100) substrate, the product was the hexagonal phase of $YMnO_3$, whereas on $LaAlO_3$(100) and STO (100) substrates, the metastable orthorhombic $YMnO_3$ phase was formed. Foncuberta's group [118] grew orthorhombic epitaxial YMO [001] (*c*-axis) and [100] (*a*-axis) textured films on conducting Nb:STO substrates. They showed a *bc*-cycloidal magnetic order exists in films of $YMnO_3$ and it can be switched to *ac*-cycloid by an appropriate magnetic field. This finding showed that a non-collinear spin arrangement in thin films allows switchable electrical polarization and tunability of their dielectric response by magnetic field; however, these experiments were carried out at 5 K. Finally, Wadati et al. [119] fabricated thin film (40 nm) of $YMnO_3$ on a $YAlO_3$ (010) substrate by PLD. They observed temperature dependent incommensurate magnetic peaks below 45 K and commensurate lattice-distortion peaks below 35 K, indicating that E-type and cycloidal states coexist below 35 K. This shows that the occurrence of the large electric polarization below 35 K is directly related to E-type magnetic ordering component in the epitaxial $YMnO_3$ films. Although these results suggest that YMO could be a suitable material for ferroelectric gate field FETs [112]; however, the high grow temperatures (≥ 800 °C) [108-118] make it impractical for integration into current applications. Further study to achieve robust ferroelectric properties and room temperature multiferroicity in manganites are pending.

**Bismuth ferrite, $BiFeO_3$ (BFO)**: The single-phase multiferroic that has been most studied in the last decade is BFO due to its multiferroic and potential magnetoelectric properties. BFO is a fully recognized room-temperature multiferroic that has all three ferroic order parameters -- ferroelectric/ferroelastic ($T_c$ = 1103 K) [120], antiferromagnetic ($T_N$ = 643 K) [121] -- and exhibits weak ferromagnetism at room temperature; however, the cross coupling among these ferroic parameters in single-phase form is very weak and poorly understood. Bulk BFO single crystal form possess a rhombohedral perovskite structure ($a = b = c$ =5.63 Å, $\alpha = \beta = \gamma = 59.4°$) [122]. In single crystal form an early report found the spontaneous polarization is ~6.1μC/$cm^2$ along the (111) direction at 77 K [123], but this is known now to have simply been a poor sample. The ferroelectricity in BFO is due to the stereo-chemically-active 6s lone pair of $Bi^{3+}$. Studies during the 1980s indicated the magnetic nature of BFO is G-type antiferromagnetic order wherein $Fe^{3+}$ ions are surrounded by six neighboring $Fe^{3+}$ ions with spin anti-parallel to the central ion [124]. The symmetry also allow a canting of the antiferromagnetic sublattices resulting in a macroscopic magnetization "weak ferromagnetism"[125]. In 2002, Palkar et al. [126] reported weak ferroelectric properties in BFO thin films deposited on $Pt/TiO_2/SiO_2/Si$ substrate, and magnetoelectric properties was demonstrated by anomaly in dielectric constant in the vicinity of Néel temperature; later, Yun et al. [127] using same substrate and technique as Palkar, but lower oxygen pressure during deposition (0.01–0.1 Torr), obtained good current–voltage characteristics and improvement in ferroelectric properties, high remanent polarization ($P_r$), $2P_r \sim 71.3$ μC/cm$^2$ and coercive field ($E_c$), $2E_c \sim 125$ kV/cm. However, in 2003 Wang et al. [25] reported epitaxial monoclinic BFO thin films grown on $SrRuO_3/STO(100)$ substrates by PLD, with a surprising of electric polarization respond up to $2P_r \sim 120$ μC/cm$^2$, which is one order of magnitude higher the value of 6.1 initially reported in from bulk BFO (an early report of high bulk polarization by Hans Schmid was unfortunately unnoticed in the literature). These films also show enhanced thickness-dependent magnetism compared to the bulk, attributed to the mismatch strain induced by heteroepitaxy. This study led to a growing interest in BFO thin films, which has endured until the present day, almost a decade after its publication. Later, Li et al. [128] investigated the effect of orientation of the STO substrate on crystallography and ferroelectrical properties of BFO. BFO films grown on (111) had a rhombohedral structure, as single crystals; whereas films grown on (101) or (001) were monoclinically distorted from the rhombohedral structure, attributed to the epitaxial constraint; the remanent polarization in each case was 55, 80, and 100 μC/cm$^2$ respectively. Afterward, a multilayer film of $BiFeO_3/La_{0.7}Sr_{0.3}MnO_3/SrTiO_3$ was deposited on Si substrate by pulsed laser deposition. The $BiFeO_3$ film showed a low leakage-current density, and a $2P_r \sim 110$ μC/cm$^2$ was observed at room temperature [129] On the other hand, Shvartsman et al. [130] showed large ferroelectric polarization ~40 μC/cm$^2$ in ceramic BFO, value was close to the theoretical predicted values. Since highest values of polarization were obtained in ceramic, polycrystalline and epitaxial films, these results indicated the larger polarization is not related to the strain but it is an intrinsic property of BFO, a result supported by first-principles calculation [131].

However device drawbacks for BFO are high leakage current, a tendency to fatigue, and thermal decomposition near the coercive field, which have been addressed by doping BFO, for example with Pr or Mn [132,133]. Due the exceptional ferroelectric properties, multiferroic nature, potential ME coupling, and its lead-free character, BFO remains a potential candidate for the next generation of ferroelectric memory applications.

**Lead-based solid solution perovskites:**

Another multiferroic single phase material family very recently studied in thin film form are lead based solid solution perovskites (SSP). To our knowledge the first publication on Pb-SSP thin films was from Ashok et al. [37]. They reported single-phase polycrystalline $(PbZr_{0.53},Ti_{0.47}O_3)_{0.8}$- $(PbFe_{0.67}W_{0.33}O_3)_{0.2}$, (0.8PZT/0.2PFW) synthesized by chemical solution deposition, a new room- temperature single phase multiferroic magnetoelectric material. Later, the multiferroic properties of single phase 0.8PZT/0.2PFW system were confirmed by Lee et al. [134,135] by growing epitaxial 0.8PZT/0.2PFW films on SRO coated STO (001) substrate. Later, $(PbZr_{0.53}Ti_{0.47}O_3)_{0.60} - (PbFe_{0.5}Ta_{0.5}O_3)_{0.4}$, (0.6PZT/0.2PFT) highly oriented thin film was synthesized by PLD on $La_{0.67}Sr_{0.33}CoO_3$ coated MgO substrates. This film shows near room temperature a frequency dependent dielectric maximum, moderate polarization and a weak magnetic moment, as well as multiferroic relaxor characteristics [136]. The synthesis of single-phase solid solution materials using conventional and relaxor ferroelectrics might open a new way to realize a room-temperature multiferroic material with strong magnetoelectric coupling.

**Multiferroic Nano-Composites: Nomenclature**

A short summary of multiferroic (ferromagnetic-ferroelectric) nano composites is given below, including multiferroic superlattices, multiferroic birelaxors with only short-range ordering of magnetization and electric polarization, and ferroelectrics with long-range ordering but short-range magnetic clustering. Although these structures can be described with the usual two-digit notation (e.g., 0-3 or 1-3, where the first digit is the dimensionality of the first phase and is chosen as less than or equal to that of the second phase, which defines the connectivity), an additional digit or subscript is necessary to discriminate among short- and long-range ordering.

The physics of composite materials is successfully categorized into classes with two-digit labels, such as 0-3, in which each digit can be from 0 to 3, the first digit describes the dimensionality of object within a matrix, and the second describes the dimensionality of its connectedness to other equivalent objects. For example, 1-3 describes a system with one-dimensional rod-like objects (such as carbon fibers) connected to each other by a three dimensional matrix, such as a polymer or epoxy. For piezoelectric composites the piezoelectric phase is conventionally designated by the first digit, even if it is the larger (e.g., 3-1 rather than 1-3), which is rarely the case [137]. This problem becomes acute only for multiferroic nano-composites, where the average size of the smaller phase can be only a few unit cells. In this case there are no grains or grain boundaries and there is a grey area between composites and relaxors. These notations do not completely specify the geometry of the system; for example, the labels do not discriminate between oriented, aligned rods within a matrix and randomly oriented rods ("PZT spaghetti").

Recently an excellent review of multiferroic composites was published [8]. The present aim is merely to suggest some convenient notation for these new families of important materials. For examples:

(i) 2-2 systems:
2-2 composites are generally the choice for tape-cast multilayer actuators, such as PZT composites. They also include most multilayer capacitors [8].
Within the class of 2-2 multiferroic composites we have examples in which both magnetic and ferroelectric ordering are long-range (Terfenol-D on PZT [138] or $BaTiO_3$ on Ni [139]) – we shall denote them and similar cases [140] as 2L-2L -- and those birelaxors such as $PbFe_{2/3}W_{1/3}O_3$ in which both magnetic and ferroelectric ordering are short-range and probably planar [18, 37,141] – which we denote as 2S-2S.

(ii) 1-3 or 3-1 systems
Most piezoelectric devices are 1-3 systems, as developed initially by Newnham and Cross and their Penn State group[137].
3-1 multiferroic composites were pioneered by developments of nano-pillars of cobalt ferrite in barium titanate [142] Here the magnetic and ferroelectric ordering is both long-range, and we denote it as 3L-1L.

(iii) 0-3 systems
Compounds such as $Pb(Fe,Ta,Zr,Ti)O_3$ [39] or $PbFe_{1/2}Ta_{1/2}O_3$ [143,24] or $PbFe_{1/2}Nb_{1/2}O_3$ [144] are nano-composites although they are nominally single-phase. This apparent paradox arises because the Fe ions cluster at the B-site and are not in random occupancy. This produces a point-like zero-dimensional object with three-dimensional connectivity through the surrounding medium. We designate this as 0S-3L, meaning that the ferroelectricity is long-range but the magnetism is short-range.

**Heterostructures composite thin films**

During the last decade, a great effort have been made to synthesize multi-phase materials thin films composed of materials that have shown large ferromagnetic and ferroelectric ordering, with the possibility of ME coupling effect. The development in multiferroic thin films has been achieved due to improved first-principles computational techniques that have aided in the design of new multiferroics and provided understanding of factors that promote coupling between magnetic and ferroelectric order parameters [1,145,146] and advances in thin film growth techniques have offered routes to fabricate structures and phases that are inaccessible by traditional chemical method [2]. The availability of high quality thin film samples in conjunction with a broad spectrum of analytical tools has improved our ability to accurately characterize multiferroic behavior and has opened the door to design of practical devices based on ME coupling. From a microstructural point view the composite thin films can classified as 0-3 particulate nanocomposite, 2-2 horizontal heterostructures and 1-3 vertical heterostructures, as was explained with more detail in the previous section.

From Table 2 we can see the most studied configuration in multiferroic magnetoelectric thin film form is 2-2 as bi-layers, multilayers, and superlattices for a diverse group of ferroelectric and magnetic materials. The popularity of 2-2 configuration is due to their ease of synthesis along with reduced leakage problem due to the blocking of current flow by resistivity ferroelectric layers. Although aside from their nature multiferroics, indirect ME coupling had been observed in different ways in 2-2 structures, evidence of direct coupling ME is absent in most of them. On the other hand, the ME values reported in thin films form are lower compared with those reported in ME laminated composites (see Table 1). Although many of the film growth techniques allow better control of the interfaces that improve the transfer the strain efficiency between two phases, the ME coupling in 2-2 configuration is affected by the a large in-plain constraint from the substrate.

Very recently a series of experiments has been carried out on bilayer samples, where the thin-film layer of ferroelectric/ferromagnetic material had been grown on the ferromagnetic/ferroelectric substrate (single crystal) film, which showed interesting magnetoelectric effects: i.e. Ni-PMN-PT, Ni presented a change in out-of-plane magnetization due to the ME effect in this heterostructure [184]; a novel approach for magnetization switching (magnetic anisotropy symmetry is altered) in Fe/BTO was observed when BTO changes from tetragonal to orthorhombic phase) [185]; large anisotropic remanent magnetization was observed in LSMO/PMN-0.3PT structures and the remanent magnetization along the [100]/[011] was suppressed/enhanced by an electric field applied to the substrate [187]. Because of the fabrication difficulty, fewer papers have been published for vertical 1-3 configurations (see Table 3); in these systems the substrate clamping is expected to be less compared with 2-2 structures. Although these configurations show multiferroic properties, no direct measurement of the ME effect is presented; the possible reason is the leakage problem due to the low resistance of the magnetic nanopillars imbedded in the ferroelectric matrix. In contrast, the ME coupling in the 0-3 structure had been measured ranging between 35-300 mV cm$^{-1}$ Oe$^{-1}$, but in most of the cases a leaky ferroelectric loop and high leakage current have been reported in these structures.

In a recent review article Nan's group [9] suggested additional confirmation is needed to determine whether the magnetic field-induced electric polarization change in fact reflects *real* ME coupling in the composite films or artifacts.

Table 3. Multiferroic heterostructures thin films.

| Structure | Materials | Technique | α (mV cm$^{-1}$ Oe$^{-1}$) | Ref. |
|---|---|---|---|---|
| Horizontal 2-2 | **2001** $La_{0.67}Sr_{0.33}MnO_x/(Pb,La)(Ca.Ti)O_3$ | MOD- ML | MR | 147 |
| | **2004** $BaTiO_3$ (BTO)/ $CoFe_2O_4$ (CFO) | Composition spread | --- | 148 |
| | $Pr_{0.85}Ca_{0.15}MnO_3/Ba_{0.6}Sr_{0.4}TiO_3$ | PLD –SL | MR | 149,150 |
| | **2005** $La_{1.2}Sr_{1.8}Mn_2O_7/PbZr_{0.3}Ti_{0.7}O_3$ | CSD | Dielectric anomaly | 151 |
| | $La_{0.7}Ca_{0.3}MnO_3/BTO$ | PLD –SL | MR, MC | 152 |
| | $BiFeO_3$ (BFO)/ $PbZr_{0.5}Ti_{0.5}O_3$ (PZT5) | CSD | --- | 153 |
| | **2006** CFO/ $PbZr_{0.52}Ti_{0.48}O_3$ (PZT) | PLD –ML | Magnetic field induces polarization. | 154 |
| | $La_{0.7}Ca_{0.3}MnO_3/BTO$ | PLD –SL | MR | 155 |
| | CFO film/BTO single crystal subs | PLD | Magnetization changes at BTO phase transitions | 156 |
| | PZT/CFO | PLD-ML | Polarization changes with magnetic field. | 157 |
| | **2007** $La_{0.6}Sr_{0.4}MnO_3/ 0.7Pb(Mg_{1/3}Nb_{2/3})O_3$-$0.3(PbTiO_3)$ | PLD-SL | Polarization changes with magnetic field. | 158 |
| | $La_{0.67}Sr_{0.33}MnO_3$ (LSMO) /BTO sub | PLD | Magnetization changes at BTO phase transitions. | 159 |
| | $BFO/Bi_{3.25}La_{0.75}Ti_3O_{12}$ | MOD-ML | --- | 160 |
| | Fe/BTO substrate | MBE | Magnetization changes at BTO phase transitions. | 161 |
| | PZT/LSMO | PLD | 4 | 162 |
| | PZT/CFO | PLD –ML | --- | 163 |
| | **2008** $PZT/SrRuO_3/CoFe_2O_4$ | PLD –ML | Polarization and dielectric change with magnetic field. | 164 |
| | PZT/CFO | Sol-gel spin-coating BL | Magnetic induced voltage change. | 165 |
| | PZT5/CFO/PZT5 | Dual cathode rf sputtering | --- | 166 |
| | $Ni_{0.23}Fe_{2.77}O/Pb(Zr,Ti)O_3$ substrate | Spin spray deposition. Combined sol-gel and rf sputtering. | Magnetization changes when applied electric field | 167 |
| | CFO/ PZT | Spin coating and rf sputtering. | 238 | 168 |
| | BFO/CFO | rf sputtering -BL | ---. | 169 |
| | PZT/CFO | PLD –ML | --- | 170 |
| | **2009** PZT/CFO | Sol-gel spin-coating BL | Magnetic field dependent Raman scattering | 171 |
| | $PZT/ PbFe_{0.66}W_{0.33}O_3$ | PLD-ML | --- | 172 |
| | PZT/CFO | PLD-ML | Magnetodielectric | 173 |
| | $Bi_2NiMnO_6$–$La_2NiMnO_6$ | PLD-SL | Magnetodielectric | 174 |
| | **2010 - present** PZT/LSMO | PLD - BL and SL | Dielectric anomaly, Polarization changes with | 175,176, 177 |
| | $Ba_{0.7}Sr_{0.3}TiO_3$(BST) /LSMO | PLD-BL and SL | magnetic field. 250 | 140,178 179 |
| | $PbZr_{0.65}Ti_{0.35}O_3$/ $La_{0.5}Sr_{0.5}CoO_3$ | MBE-magnetron Sputtering | --- | 180 |
| | $PbZr_{0.2}Ti_{0.8}O_3$/ $La_{0.8}Sr_{0.2}MnO_3$ | | Spin control by electric field. | 181 |
| | NiFe/BTO/SRO | PLD-BL | Electric field manipulated magnetization | 182 |
| | $BiFeO_3$/CoFe | CSD- sputtering | --- | 183 |
| | Ni/[Pb($Mn_{1/3}Nb_{2/3}$)$O_3$]$_{0.68}$-[$PbTiO_3$]$_{0.32}$ single crystal | E-beam evaporated (Ni) | Magnetoelectric manipulation of domain wall configuration. | 184 |
| | Fe/$BaTiO_3$ single crystal | MBE | Magnetization switching via interface lattice distortion. | 185 |
| | Ni/PZT | Sol gel-PZT and magnetron sputter-Ni | Electric voltage induced magnetization. | 186 |
| | LSMO/0.7Pb($Mg_{2/3}Nb_{1/3}$)$O_3$-0.3$PbTiO_3$ single crystal | rf sputtering | Large anisotropic remanent magnetization tunability. | 187 |
| Vertical 1-3 | **2004-2012** BTO/CFO | PLD | Magnetization changes at BTO phase transitions. | 142,188, 189 |
| | CFO/$PbTiO_3$ | PLD | ---- | 190 |
| | BFO/CFO | PLD-combinatorial method | ---- | 191,192 |
| Particulate Nanocomposite 0-3 | **2003-2008** $Ba_{0.5}Sr_{0.5}TiO_3$ /$BaFe_{12}O_{19}$ | rf magnetron sputtering | --- | 193 |
| | CFO/PZT | Sol-gel process | 317 | 194 |
| | BTO/CFO | PLD | --- | 195 |
| | $Bi_{3.15}Nd_{0.85}Ti_3O_{12}$–$CoFe_2O_4$ | CSD | 35 | 196 |
| | BTO-Co | rf magnetron sputtering | 160-170 | 197 |
| | PZT/CFO | PLD | Magneto-capacitance | 198 |

Metal organic deposition:MOD, Chemical solution deposition:CSD, Pulsed laser deposition:PLD, Molecular beam epitaxy (MBE), Magnetoresistance:MR, Magnetocapacitance:MC, Bilayer:BL, Multilayer:ML, superlattices:SL

**Magnetic control of Ferroelectric Polarization in PZT/LSMO Bilayer Structures**

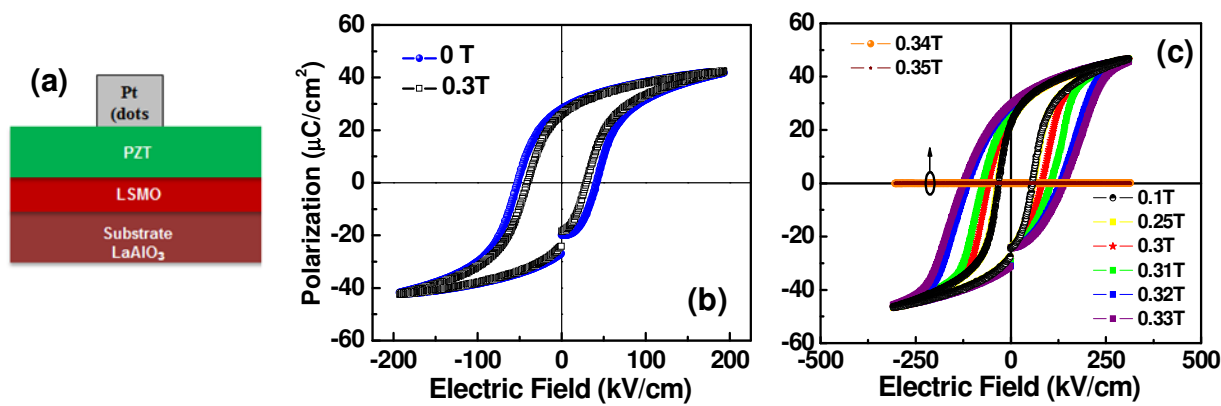

Figure 5. (a) A sketch of the Pt/PZT/LSMO bilayer capacitor. Ferroelectric hysteresis loop of PZT films with LSMO electrodes on $LaAlO_3$ substrates at different applied magnetic fields for: (b) PZT film thickness 1.4 μm. No significant magnetic field dependence is observed. (c) PZT film thickness 0.55 μm. A significant magnetic field dependence is observed in the hysteretic loss near a critical field of $H$ = 0.34 T. Adapted from Ref. [177], reproduced with permission (© 2011 IOP).

In an attempt to make artificial magnetoelectric multiferroics heterostructures, Dussan et al. synthesized Pb(Zr$_{0.53}$Ti$_{0.47}$)O$_3$/La$_{2/3}$Sr$_{1/3}$MnO$_3$ (PZT/LSMO) bilayers thin films on LaAlO$_3$(100) (LAO) substrates using pulsed-laser deposition (PLD); electrical measurements were carried out on metal–insulator–metal capacitor structures fabricated with a Pt top electrode diameter of 200 µm deposited by dc sputtering through a shadow mask, (see Fig. 5 (a)). The films have multiferroic behavior. Additionally, they discovered that the ferroelectric polarization is suppressed entirely with the application of a magnetic field of merely 0.34 T. This effect occurred when the PZT layer thickness was ~ 550 nm (Figure 5 (c)), the hysteresis curve broadens remarkably width increase of applied magnetic field (*H*), and then vanishes completely at *H* = 0.34 T. This effect was completely reversible; and at room temperature. But it did not happen in the samples with PZT thickness of around 1.4 µm (Figure 5 (b)). This result was explained in the light of the negative magnetoresistance effect of the LSMO, which became smaller in the applied *H* causing the increases the current flow and hence the dielectric loss in the capacitor. Since the modest magnetic field (0.34 Tesla) quenched the ferroelectric polarization (+*P* or –*P* to zero), thus generating strong interest in its usage in various sensors and electronics applications [175-177].

**Strong Magnetoelectric effect in Ba$_{0.7}$Sr$_{0.3}$TiO$_3$ and La$_{0.7}$Sr$_{0.3}$MnO$_3$ (BST/LSMO) Superlattices**

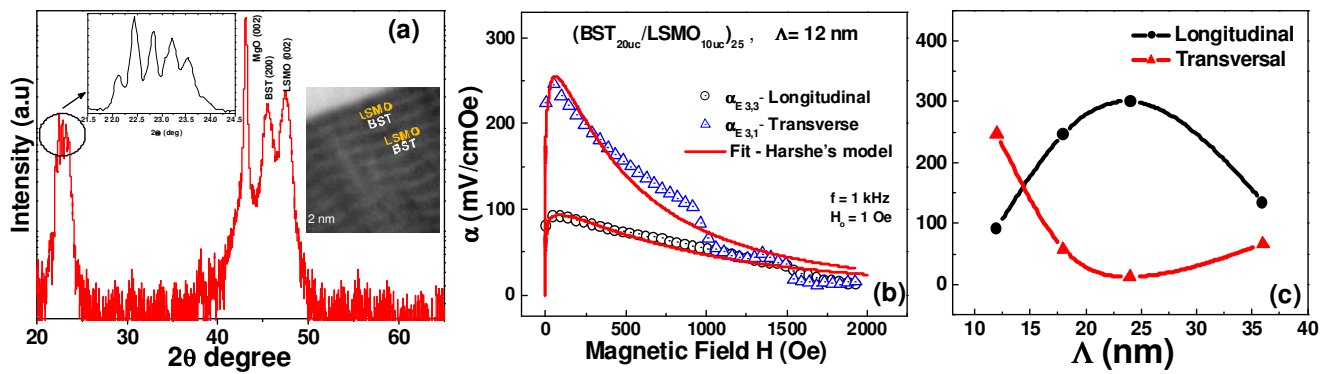

Figure 6. (a) X-ray diffraction pattern of a BST$_{20uc}$/LSMO$_{10uc}$ SL on MgO substrate. The upper inset show the satellite peak around the (001) peak, lower inset show the transition electron microscope (TEM) that evidence of the formation of superlattices structure. Longitudinal and transverse modes of magnetoelectric voltage coefficient ($\alpha$) at 1 kHz and room temperature (b) as a function of the bias magnetic field for (BST$_{20uc}$/LSMO$_{10uc}$)$_{25}$ superlattices. Solid lines represent the fitting using the Harshe model. (c) as a function of the superlattices periodicity. Adapted from Ref. [179], reproduced with permission (© 2012 AIP).

Martinez et al. [140, 178-179] investigated the structural analysis and multiferroic and magnetoelectric properties of Ba$_{0.7}$Sr$_{0.3}$TiO$_3$/La$_{0.7}$Sr$_{0.3}$MnO$_3$ superlattice (SL) thin films grown on LaNiO$_3$ (LNO) coated (100) MgO substrate by pulsed laser deposition. The thickness of the superlattice was keep constant (300 nm) irrespective of the modulation period in different SL-configurations. The x-ray diffraction along with cross-sectional TEM of the (BST$_{20uc}$/LSMO$_{10uc}$)$_{25}$ show the SL structure was present in the thin films (Figure 1(a)). A manuscript with a detailed study of the interface effect using high resolution TEM is in preparation. The direct ME measurement illustrates a strong response of superlattices films,

Figure 6(b) show the typical ME response of one of these SLs: a maximum peak for $\alpha_{E,33}$ and $\alpha_{E,31}$ at ~250 Oe and ~90 Oe respectively was seen with further monotonically decrease to zero with increase in bias magnetic field (H). A similar H-dependent behavior of the ME coefficients was seen for the superlattices with other periodicities (i.e, $\Lambda$=18, 24 and 36 nm). The experimental data fitted well with a modified free-body model, (solid lines in Figure 6(b)). From Figure 6(c), the ME values follow a hyperbolic behavior which increases for the longitudinal ($\alpha_{E33}$), until $\Lambda$=24 nm with further decrease in periodicities. A similar but inverse character is present in the case of the transverse ME coupling ($\alpha_{E31}$). However, the evolution of the ME coefficient as a function of periodicity requires further research on interface-microstructure-property relations; at present it is very poorly understood. These results showed the substrate clamping and the resulting weakening of ME coupling that are usually encountered in piezoelectric-magnetostrictive bilayers was not observed in the systems studied here. The ME voltage coefficients are in the range 35-300 mV/cm Oe is as high as measured values in thick film and bulk ME composites.

**Device Functionality Considerations**

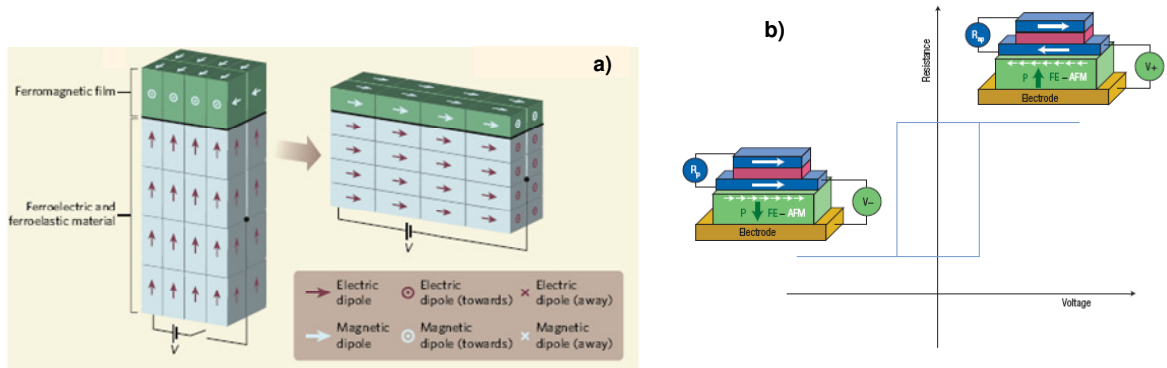

Figure 7. (a) Electrical control of ferromagnetic films. Adapted from Ref. [199], reproduced with permission (© 2008 NPG). (b) Sketch of a possible magnetoelectric random access memory (MERAM). Adapted from Ref. [200], reproduced with permission (© 2008 NPG).

Multiferroic ME materials offer the opportunity to study novel physical phenomena and they present the possibility for its application in new multi-functional devices [200]; for example, as transducers converting between magnetic and electric fields, attenuators, filters, field probes and data recording devices based on electric control of magnetization and vice versa [201]. With the coexistence of several order parameters in multiferroics and the ME coupling can both be exploited in novel types of memory [9]. As ferroelectric polarization and magnetization are used to encode binary information in FeRAM (ferroelectric random access memory), and MRAM (magnetic random access memory) respectively, the coexistence of magnetization and polarization in a multiferroic material allows the realization of four-state logic states (or >4) in a single device. The control of magnetic domains in a ferromagnetic film through an adjacent voltage-driven multiferroic promises to combine the best features of FeRAM and MRAM. However this promise is very distant, with many challenges beyond the unequivocal at buried and imperfect interfaces. Recently N. Mathur [199] proposed the creation and study of a single-domain multiferroic crystal, because multiferroic materials have complex domain structures that make precise interpretation of their behavior difficult. Figure 7 (a) illustrates the electrical control of ferromagnetism in multilayer films of the ferroelectric–ferroelastic and ferromagnetic sublayers. In such multilayer structures an electrically driven change in the electrical polarization of a ferroelectric-ferroelastic layer produces mechanical strain, which is transmitted to an overlaying ferromagnetic thin film. This deformation modifies the preferred orientations of the magnetic dipoles and therefore the macroscopic magnetizations. Bibes et al. [200] explained the basic operation of magnetoelectric random access memory (MERAM). Figure 7 (b) shows the binary information is stored by the magnetization direction of the bottom ferromagnetic layer (blue), read by the resistance of the magnetic trilayer ($R_p$ when the magnetizations of the two ferromagnetic layers are parallel), and written by applying a voltage across the multiferroic ferroelectric–antiferromagnetic layer (FE-AFM; green). If the magnetization of the bottom ferromagnetic layer is coupled to the spins in the multiferroic (small white arrows) and if the magnetoelectric coupling is strong enough, reversing the ferroelectric polarization $P$ in the multiferroic changes the magnetic configuration in the trilayer from parallel to antiparallel, and the resistance from $R_p$ to antiparallel ($R_{ap}$). A hysteretic dependence of the device resistance with voltage is achieved (blue curve).

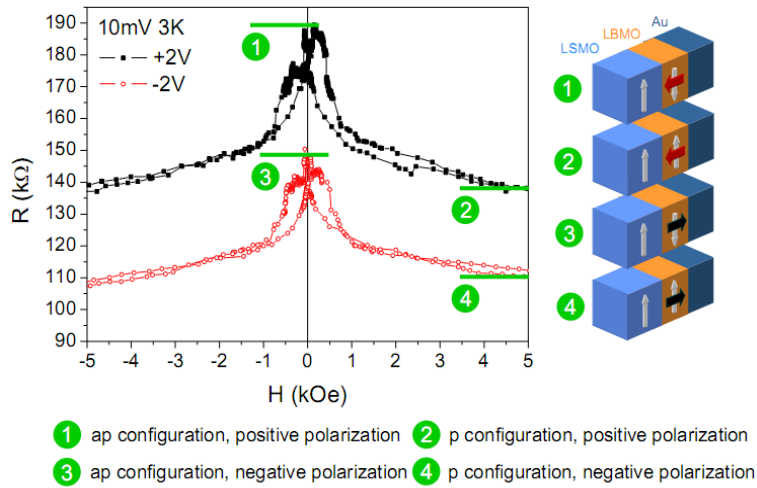

Figure 8. (a) Tunnel magnetoresistance properties of a LSMO/LBMO(2nm)/Au junction. In the legends the meaning of ap:anti-parallel and p:parallel. Adapted from Ref. [205], reproduced with permission (© 2008 NPG).

Experimental and theoretical findings suggest that FE persist in ultrathin films (~3 nm) [202], which open possibility to further miniaturize devices based on FE materials. A way to exploit the FE properties of ultrathin films and multifunctional character of MF materials is to design tunnel junctions (TJ) integrating FE or MF films as the tunnel barrier or making TJ itself a multiferroic heterostructures. At present TJ using insulator (nonpolar) barrier is the base of MRAMs [203]. The FE and MF heterostructures mentioned above may take advantage of tunnel magnetoresistance (TMR) and tunnel electroresistance (TER) effects to obtain a valuable room temperature four-resistance-state based material. It will have a broader impact due to the potential applications of TJ for the fabrication of devices, such as memory devices. Recent theoretical and experimental studies have demonstrated the existence of TMR and TER properties in TJ, using polar barrier (ferroelectric/multiferroic tunnel junctions (FTJ/MFTJ)) [204]. Several studies have been performed using a multiferroic as the barrier material. Gajek et al. [205, 206] have fabricated LSMO/La$_{0.1}$Bi$_{0.9}$MnO$_3$ (LBMO) (2 nm)/Au hetero-junctions. (LBMO is a well known multiferroic at low temperature) and found the signature of four resistive states with tunnel magnetoresistance curves at 4 K at V$_{d.c.}$ = 10mV, after applying a voltage of ±2 V. However, this result has been obtained experimentally at very low temperature (< 10 K), and is not suitable for device applications at room temperature.

**Conclusions**

The last fifteen years, a continuous interest has been evolved in the research of magnetoelectric multiferroic materials by the scientific community not only in the search of new materials and structures with ME multiferroic properties with high direct ME response but also in the implementation the new experimental techniques that allow one to visualize changes in the magnetic response of multiferroic material due to the application of an electrical signal and vice versa, plus the use of this response in practical applications in multifunctional devices such as sensors, transducers, memories and spintronics.

Recent experiments have demonstrated the potential applications of ME materials in prototype devices; hopefully one or more should come onto the market in near future. Involvement of a large number of scientific groups and exponential growth in the number of publications in this area indicate their potential functionality as materials and devices.

**Acknowledgments**

This work was partially financially supported by DoD W911MF-11-1-0204, IFN-NSF-RII 07-01-25 and DoE-DE-FG02-08ER46526 grants.

**References**

[1] N. A. Hill, J. Phys. Chem. B 104, 6694 (2000).
[2] R. Ramesh and N.A. Spaldin. Multiferroics: progress and prospects in thin films. Nature Materials. 6, 21 (2007).
[3] J. Busch-Vishniac, *Phys. today*, **51**, 28 (1998).
[4] N. Fujimura, T. Ishida. T. Yoshimura and T. Ito, *Appl. Phys. Lett.* **69**, 1011 (1996).
[5] V.E. Wood and A.E. Austin, in *Magnetoelectric interaction phenomena in crystals,* eds. A. J. H. Schmid, Gordon and Breahc, 1975.
[6] N.A. Spaldin, W.E.Pickett. Computational design of multifunctional. J. Solid sta. Chem. 176, 615-632 (2003).
[7] W. Prellier, M.P. Singh, P. Murugavel. The single-phase multiferroic oxides: from bulk to thin film. J. Phys.: condens. Matter. **17** R803-R832. (2005)
[8] C-W Nan, M-I Bichurin, S. Dong, D. Viehland and G. Srinivasan. Multiferroic magnetoelectric composites: Historical perspective, status, and future directions. *J. Appl. Phys*. 103 031101 (35pp) (2008)
[9] Y.Wang. J. Hu, Y. Lin, C.W. Nan. Multiferroic magnetoelectric composite nanostructure. NPG Asia Mater. 2 61-68 (2010).
[10] W.C Rontgen. *Ann. Phys*. **35** 264 (1888).
[11] H.A Wilson. *Phil. Trans R. Soc*. A. **204** 129 1905.

[12] I. E. Dzyaloshinskii. On the magneto-electrical effects in antiferromagnetics. Soviet Physics JETP, 37 628-629. (1960)

[13] D.N. Astrov, *JETP (UURS).* **38**, 984 (1960).

[14] D.N. Astrov, *Sov. Phys. JETP.* **13**, 729 (1961).

[15] V.J. Folen, G.T. Rado, and E.W. Stalder. *Phys. Rev. Lett.* **6**, 607 (1961).

[16] V G.T. Rado and J. Folen. *Phys. Rev. Lett.* **7**, 310 (1961).

[17] H Schmid, Multi-ferroic magnetoelectric. *Ferroelectrics.* **162** 317-338. (1994)

[18] Pirc R, Blinc R and Scott JF: Mesoscopic model of a system possessing both relaxor ferroelectric and relaxor ferromagnetic properties. *Phys. Rev. B.* **79:** 214114-7p. (2009)

[19] Nomura S, Takabayashi H, Nakagawa T: Dielectric and magnetic properties of $Pb(Fe_{1/2}Ta_{1/2})O_3$. *Jpn. J. Appl. Phys.* **7**: 600-604. (1968).

[20] Brixel W, Rivera JP, Steiner A and Schmid H: Magnetic field induced magnetoelectric effects, (ME)H, in perovskite $Pb_2Co_2WO_6$ and $Pb_2FeTaO_6$. *Ferroelectrics.* ;**79:** 201-204. (1988)

[21] Lampis N, Franchini C, Satta G, Geddo-Lehmann G and Massidda S: Electronic structure of $PbFe_{1/2}Ta_{1/2}O_3$: Crystallographic ordering and magnetic properties. *Phys. Rev. B.;* **69:** 064412-12p. (2004)

[22] Palkar VR and Malik SK: Observation of magnetoelectric behavior at room temperature in $Pb(Fe_xTa_{1-x})O_3$. *Solid State Commun.* **134:** 783-786. (2005)

[23] S. B. Majumder, S. Bhattacharya, R. S. Katiyar, A. Manivannan, P. Dutta, and M. S. Seehra, *J. Appl. Phys.* **99**, 024108 (2006)

[24] R. Martinez, R. Palai, H. Huhtinen, J. Liu, J. F. Scott, and R. S. Katiyar. Nanoscale ordering and multiferroic behavior in $Pb(Fe_{1/2}Ta_{1/2})O_3$. *Phys. Rev. B* **82**, 134104 (2010).

[25] Wang J, Neaton JB, Zheng H, Nagarajan V, Ogale SB, Liu B, Viehland D, Vaithyanathan V, Schlom DG, Waghmare UV, Spaldin NA, Rabe KM, Wuttig M, Ramesh R: Epitaxial $BiFeO_3$ multiferroic thin film heterostructures. *Science.* **299:** 1719-1722. (2003)

[26] W. Eerenstein, N. D. Mathur, J. F. Scott, Multiferroic and magnetoelectric materials, Nature, 442, 759-764. (2006)

[27] W. Eerenstein, F. D. Morrison, J. Dho, M. G. Blamire, J. F. Scott, and N. D. Mathur, "Comment on Epitaxial $BiFeO_3$ multiferroic thin film heterostructures," *Science*, **419**, 1203a, (2005).

[28] Y.H. Chu, L.W. Martin, M.B. Holcomb, R. Ramesh. Controlling magnetism with multiferroics. Materials Today. 10 16-23 (2007)

[29] T. Kimura, T. Goto, H. Shintani, K. Ishizaka, T. Arima, Y. Tokura. Magnetic control of ferroelectric polarization. Nature 426, 55-58 ( 2003)

[30] B.B Nan Aken,. B.B. Meetsma, T.T.M. Palstra. Hexagonal $LuMnO_3$ revisited. Acta cryst. E57, 101-103 (2001).

[31] S.C. Abrahams. Ferroelectricity and structure in the $YMnO_3$ family. Acta cryst. B57, 485 (2001).

[32] R. Kajimoto, H. Yoshizawa, H. Shintani, T. Kimura, Y. Tokura, Magnetic structure of $TbMnO_3$ by neutron diffraction. Phys. Rev. B. 70, 012401 (2004).

[33] G.A. Smolensky, V. Aisupov, A.I. Agronovskaya, New ferroelectric of complex composition of the type $(A_2^{2+}(B_I^{3+} B_{II}^{5+})O_6$. Sov. Phys. Solid State. 1 150 (1959).

[34] A. Kumar, I. Rivera, R. S. Katiyar, and J. F. Scott. Multiferroic Pb(Fe$_{0.66}$W$_{0.33}$)$_{0.80}$Ti$_{0.20}$O$_3$ thin films: A room-temperature relaxor ferroelectric and weak ferromagnetic. *Appl. Phys. Lett*, **92**, 132913, (2008).

[35] B.-J. Fang, C.-L. Ding, W. Liu, L.-Q. Li, L. Tang. Preparation and electrical properties of high-Curie temperature ferroelectrics. *Eur. Phys. J. Appl. Phys*. 45, 20302, (2009).

[36] M. Yokosuka, *Jpn. J. Appl. Phys*. Electrical, Electromechanical and Structural Studies on Solid Solution Ceramic Pb(Fe$_{1/2}$Nb$_{1/2}$)O$_3$–Pb(Zn$_{1/3}$Nb$_{2/3}$)O$_3$. 38, 5488 (1999).

[37] A. Kumar, G. L. Sharma, R.S. Katiyar, J. F. Scott, R. Pirc, R. Blinc, Magnetic control of large room-temperature polarization, J. Phys. Condens. Matter. 21 (2009) 382204 (7 pp).

[38] L. Mitoseriu, D. Marré, A.S. Siri, A. Stancu, C.E, Fedor, P. Nammi. Magnetoelectric coupling in the multiferroic Pb(Fe$_{0.66}$W$_{0.33}$)O$_3$-PbTiO$_3$ system. J. Opt. Adv. Mater. 6, 723-728 (2004)

[39] D. A. Sanchez, N. Ortega, A. Kumar, R. Roque-Malherbe, R. Polanco, J. F. Scott and R. S. Katiyar. Symmetries and Multiferroic Properties of New Room-Temperature Magnetoelectrics: Lead iron tantalate – lead zirconate titanate (PFT/PZT). AIP Advances, 1 (2011) 042169 (14pp).

[40] B. M. Fraygola, A. A. Coelho, D. Garcia, J. A. Eiras. Magnetic and Dielectric Proprieties of Multiferroic (1-x)Pb(Fe$_{2/3}$W$_{1/3}$)O$_3$-xPbTiO$_3$ Ceramics Prepared Via a Modified Two-stage Solid-state Reaction. Materials Research. 14 434-441 (2011).

[41] Y. Koyata, H. Nakamura, N. Iwata, A. Inomata and K. Kohn. Electric and magnetic low-temperature phase transitions of YbMn$_2$O$_5$ *J. Phys. Soc. Japan* **65** 1383 (1996)

[42] Koyata Y and Kohn K Low-temperature phase transition in ErMn$_2$O$_5$ *Ferroelectrics* **204** 115 (1997).

[43] N. Iwata, M. Uga, and K. Kohn. Magnetic-field-induced transition of thulium manganese oxide TmMn$_2$O$_5$. *Ferroelectrics* **204** 97 (1997).

[44] M. Uga, N. Iwata, and K. Kohn Magnetoelectric effect of TmMn$_2$O$_5$. *Ferroelectrics* **219** 55 (1998).

[45] Popov Y F, Kadomtseva A M, Krotov S S, Vorob'ev G P, Kamilov K I, Lukina M M and TegranchiM M Magnetic and structural phase transitions in YMn$_2$O$_5$ ferromagnetoelectric crystals induced by a strong magnetic field *J. Exp. Theor. Phys.* **96** 961 (2003)

[46] Hur N, Park S, Sharma P A, Guha S and Cheong S-W Colossal magnetodielectric effects in DyMn$_2$O$_5$ *Phys. Rev. Lett.* **93** 107207 (2004).

[47] S-W Cheong, M. Mostovoy. Multiferroics: a magnetic twist for ferroelectricity. Nature Materials 6 13-20 (2007).

[48] M. Mercier. Doctoral Thesis Faculty of Science, University if Grenoble, France (1969).

[49] D. Vaknin, J.L. Zarestky, L.L Miller, J.P. Rivera and H.Schmid. Weakly coupled antiferromagnetic planes in single-crystal LiCoPO$_4$. *Phys. Rev. B*. 65 224414 (2002).

[50] E. Ascher, H. Schmid, D. Tar. Solid. Dielectric properties of boracites and evidence for ferroelectricity. Sate Commun. **2** 45 (1964).

[51] Y.Y. Tomashpol'ski, Y.N. Venevtsev, V.N. Beznozdrev. Fiz. Tverd. Tela 7 2763 (1965)

[52] H. Schmid, H. Rieder, E. Ascher. Magnetic susceptibilities of some 3*d* transition metal boracites Solid. Sate Commun. 3 327 (1965).

[53] E. Ascher, H. Rieder, H. Schmid and H. Stössel. Some Properties of Ferromagnetoelectric Nickel-Iodine Boracite, Ni$_3$B$_7$O$_{13}$I. *J. Appl. Phys*. 37, 1404 (1966).

[54] M. N. Iliev, V. G. Hadjiev, M. E. Mendoza, J. Pascual. Raman spectroscopy of multiferroic trigonal boracite Co$_3$B$_7$O$_{13}$Cl. *Phys. Rev. B* 76, 214112 (2007).

[55] D.L. Fox, J.F. Scott. Ferroelectrically induced ferromagnetism. J. Phys. C 10 L329-L331 (1977).

[56] J. F. Scott. Mechanism of dielectric anomalies in BaMnF$_4$. Phys. Rev. B. 16, 2329-2331 (1977).

[57] J. F Scott. and D. R. Tilley. Magnetoelectric anomalies in BaMnF$_4$. *Ferroelectrics* **161** 235–43 (1994).

[58] C. Ederer and N. A. Spaldin. Electric-field-switchable magnets: the case of BaNiF$_4$. *Phys. Rev.* B **74** 020401 (2006).

[59] C. Ederer and N. A. Spaldin Origin of ferroelectricity in the multiferroic barium fluorides BaMF$_4$: a first principles study *Phys. Rev.* B **74** 024102 (2006)

[60] H. Schmid. Magnetoelectric Interaction Phenomena in Crystal; A.J. Freeman and H. Schmid Eds. Gordon and breach. Newark, NJ, 1975.

[61] P.K. Baltzer. H.W. Lehman, M. Robbinson. Insulating ferromagnetic spinels. Phys. Rev. Lett. 15, 493-495 (1965).

[62] J. Hemberger, P. Lunkenheimer, R. Fichtl, H.-A. Krug von Nidda, V. Tsurkan, A. Loidl Relaxor ferroelectricity and colossal magnetocapacitive coupling in ferromagnetic CdCr$_2$S$_4$. Nature 434, 364-367 (2005).

[63] G. Catalan. Magnetocapacitance without magnetoelectric coupling. *Appl. Phys. Lett.* **88** 102902 (2006).

[64] T Rudolf, C Kant, F Mayr, J Hemberger, V Tsurkan and A Loidl. Spin–phonon coupling in ZnCr$_2$Se$_4$. *Phys.Rev.* B **75** 052410 (2007).

[65] G. Catalan and J.F. Scott. Is CdCr$_2$S$_4$ a multiferroic relaxor? *Nature* **448** E4 (2007).

[66] S Crohns, S Shrettle, P Lunkenheimer and A. Loidl. Colossal magnetocapacitive effect in differently synthesized and doped CdCr$_2$S$_4$. *Physica* B **403** 4224 A (2009).

[67] T Kimura, J C Lashley and A P Ramirez. Inversion-symmetry breaking in the noncollinear magnetic phase of the triangular-lattice antiferromagnet CuFeO$_2$ *Phys. Rev.* B **73** 220401(R) (2006).

[68] Quirion G, Plumer M L, Petrenko O A, Balakrishnan G and Proust C Magnetic phase diagram of magnetoelectric CuFeO$_2$ in high magnetic fields. *Phys. Rev.* B **80** 064420 (2009).

[69] S Seki, Y Onose and Y Tokura. Spin-driven ferroelectricity in triangular lattice antiferromagnets ACrO$_2$ (A = Cu, Ag, Li, or Na). *Phys. Rev. Lett.* **101** 067204 (2008).

[70] M. Fiebig. Revival of the magnetoelectric effect. J. Phys. D: Appl. Phys. 38 R123-R152 (2005)

[71] N.A. Spaldin, S.W. Cheong, R. Ramesh. Multiferroic: Past, present, and future. Phys. Today. 38-43 (2010).

[72] S. Picozzi, C. Ederer. First principles studies of multiferroic materials. J. Phys.: condens. Matter. 21 303201 (18pp). (2009)

[73] C.W. Nan. Magnetoelectric Effect in composites of piezoelectric and piezomagnetic phases. Phys. Rev. B. 50 6082-6088 (1994).

[74] A.M J.G Van Run, D.R Terrell, and J.H Scholing. An *in situ* grown eutectic magnetoelectric composite material. *J. mater Sci*. **9,** 1710 (1974).

[75] J. Van Suchetelene, *Philips Research Report*. **27**, 28 (1972).

[76] J. Boomgaard, and R.A.J. Born. A sintered magnetoelectric composite material BaTiO$_3$-Ni(Co, Mn) Fe$_2$O$_4$. *J. Mater Sci*. **13**, 1538 (1978).

[77] J. Van den Boomgaard, Van Run A.M J.G, and Suchetelene J. Van. *Ferroelectrics.* **10**, 295 (1976).

[78] R.P. Mahajan, K. K. Patankar, M.B. Kothale and S.A. Patil. *Bull. Mater. Sci.* 23, 273 (2000).

[79] Y.R. Dai, P. Bao, J.S. Zhu, J.G. Wan, H.M. Shen and J.M. Liu. Internal friction study on CuFe$_2$O$_4$/PbZr$_{0.53}$Ti$_{0.47}$O$_3$ composites. *J. Appl. Phys.* 96 5687 (2004).

[80] R.P. Mahajan, K. K. Patankar, M.B. Kothale, S.C. Chaudari, V.L. Mathe, and S.A. Patil. Magnetoelectric effect in cobalt ferrite–barium titanate composites and their electrical properties. *Pramana- J. Phys.* 58 1115 (2002)

[81] L. Weng, Y. Fu, S. Song, J. Tang, and J. Li. Synthesis of lead zirconate titanate–cobalt ferrite magnetoelectric particulate composites via an ethylenediaminetetra acetic acid–citrate gel process. *Scripta Materialia*. **56**, 465 (2007).

[82] C. Miclea, C. Tanasoiu, L. Amarande, C. F. Miclea, C. Plavitu, M. Cioangher, L. Trupina, C. T. Miclea, T. Tanasoiu, M. Susu, I. Voicu, V. Malczanek, A. Ivanov, C. David. Magnetoelectric properties of multiferroic cobalt ferrite and soft piezoelectric PZT particulate composites. *J Optoelectron Adv. M*. 12, p. 272 – 276 (2010).

[83] J. Zhai, N. Cai, Z. Shi, Y. Lin and C-W Nan. Magnetic-dielectric properties of NiFe$_2$O$_4$/PZT particulate composites *J. Phys. D: Appl. Phys.* **37,** 823 (2004).

[84] J. Ryu, A.V. Carazo, K. Uchino, and H.E. Kim. Piezoelectric and magnetoelectric properties of lead zirconate titanate/ Ni-ferrite particulate composites. *J. Electroceram*. **7**, 17 (2001).

[85] N.G. Kim, Y.S. Koo, C.J. Won, N. Hur, J.H. Jung, J. Yoon, Y. Jo, and H. Jung. *J.* Magnetodielectric effect in BaTiO$_3$–LaMnO$_3$ composites. *Appl. Phys.* 1**02** 014107-1 (2007).

[86] K. Mori, and M. Wuttig. Magnetoelectric coupling in Terfenol-D/polyvinylidenedifluoride composites. *Appl. Phys Lett*. **81** 100 (2002).

[87] R.S. Devan and B.K. Chougule. Effect of composition on coupled electric, magnetic, and dielectric properties of two phase particulate magnetoelectric composite *J. Appl. Phys*. **101** 014109-1 (2007).

[88] X.M. Chen, Y.H. Tang, I-W Chen, Z.C. Xu, and Y. Wu. Dielectric and Magnetoelectric Characterization of CoFe$_2$O$_4$/Sr$_{0.5}$Ba$_{0.5}$Nb$_2$O$_6$ Composites. *J. Appl. Phys*. **96** 6520 (2004).

[89] P. Dey, T.K. Nath, M.L. Nanda Goswami and T.K. Kundu. Room temperature ferroelectric and ferromagnetic properties of multiferroics xLa$_{0.7}$Sr$_{0.3}$MnO$_3$–(1−x)ErMnO$_3$ (weight percent x = 0.1, 0.2) composites. *Appl. Phys Lett*. **90** 162510 (2007).

[90] X-M Liu, S-Y Fu,Ch-J Huang. Synthesis and magnetic characterization of novel CoFe$_2$O$_4$–BiFeO$_3$ nanocomposites. *Materials Science and Engineering B* **121** 255–260. (2005).

[91] M. V. Ramanaa, N. R. Reddya, G. Sreenivasulu, K.S Kumar, B.S. Murty, V.R.K. Murthy. Enhanced mangnetoelectric voltage in multiferroic particulate Ni$_{0.83}$Co$_{0.15}$Cu$_{0.02}$Fe$_{1.9}$O$_{4-\delta}$/ PbZr$_{0.52}$Ti$_{0.48}$O$_3$ composites – dielectric, piezoelectric and magnetic properties. *Current Applied Physics* **9** 1134–1139, (2009).

[92] J. Ryu, Ch-W Baek, N-K Oh, G. Han, J-W Kim, B-D Hahn,W-H Yoon, D-S Park, J-J Kim, D-Y Jeong. Effect of Microstructure on Magnetoelectric Properties of 0.9Pb(Zr$_{0.52}$Ti$_{0.48}$)O$_3$-0.1Pb(Zn$_{1/3}$Nb$_{2/3}$)O$_3$ and Ni$_{0.8}$Zn$_{0.2}$Fe$_2$O$_4$ Particulate Composites. *Jap. J. App. Phys*. **50** 111501 (2011).

[93] J. Zhai, N. Cai, Z. Shi, Y. Lin, and C-W Nan. Coupled magnetodielectric properties of laminated PbZr$_{0.53}$Ti$_{0.47}$O$_3$ -NiFe$_2$O$_4$ Ceramics. *J. Appl. Phys.* **95**, 5685, (2004).

[94] G. Srinivasan, E.T. Rasmussen, J. Gallegos, R. Srinivasan, Y.I. Bokhan and V.M. Laletin. Magnetic bilayers and multilayer structures of magnetostrictive and piezoelectric oxides. *Phys. Rev. B*. **64** 214408 (2001).

[95] G. Srinivasan, E.T. Rasmussen, A.A. Bush, K.E. Kamentsev, V.F. Meshcheryakov and Y.K. Fetisov. Structural and magnetoelectric properties of $MFe_2O_4$–PZT (M=Ni,Co) and $La_x(Ca,Sr)_{1-x}MnO_3$–PZT multilayer composites. *Appl. Phys. A*. **78**, 721 (2004).

[96] J-P Zhou, H-C He, Z-Shi, G. Liu, and C-W Nan, Dielectric, magnetic, and magnetoelectric properties of laminated $PbZr_{0.52}Ti_{0.48}O_3/CoFe_2O_4$ composite ceramics. *J. Appl. Phys.* **100**, 094106 (2006).

[97] G. Srinivasan, E.T. Rasmussen, B.J. Levin, R. Hayes. Magnetoelectric effects in bilayers and multilayers of magnetostrictive and piezoelectric perovskite oxides. *Phys. Rev. B*. **65** 134402 (2002).

[98] N. Cai, J. Zhai, C. W Nan, Y. Lin and Z. Shi. Dielectric, ferroelectric, magnetic, and magnetoelectric properties of multiferroic laminated composites. *Phys. Rev. B*. **68** 224103 (2003).

[99] N. Cai, C.W Nan, J. Zhai, and Y. Lin. Large high-frequency magnetoelectric response in laminated composites of piezoelectric ceramics, rare-earth iron alloys and polymer. *Appl. Phys. Lett.* **84** 3516 (2004).

[100] Y. Lin, N. Cai, J. Zhai, G. Liu, and C-W. Large high-frequency magnetoelectric response in laminated composites of piezoelectric ceramics, rare-earth iron alloys and polymer. *Phys. Rev. B*. **72**, 012405 (2005).

[101] Z. Shi, W. Nan, J. Zhang, N. Cai, J-F Li. Magnetoelectric effect of $Pb(Zr,Ti)O_3$ rod arrays in a $(Tb,Dy)Fe_2$/epoxy medium. *Appl. Phys. Lett.* **87** 012503 (2005).

[102] J. Ryu, A. Vásquez Carazo, K. Uchino, and H.E Kim. Magnetoelectric properties in piezoelectric and magnetostrictive laminate composite. *Jpn J. Appl. Phys.* **40** 4948 (2001).

[103] S. G. Lu, J. Z. Jin, X. Zhou, Z. Fang, Q. Wang, Q. M. Zhang. Large magnetoelectric coupling coefficient in poly(vinylidene fluoride-hexafluoropropylene)/Metglas laminates *J. App. Phys* 110, 104103 (2011).

[104] CW Nan. Magnetoelectric effect in composites of piezoelectric and piezomagnetic phases *Phys. Rev. B* **50**, 6082–6088 (1994).

[105] C. L. Zhang, J. S. Yang, and W. Q. Chen. Harvesting magnetic energy using extensional vibration of laminated magnetoelectric plates. *Appl. Phys. Lett.* **95**, 013511 (2009).

[106] C. L. Zhang and W. Q. Chen. Magnetoelectric coupling in multiferroic laminated plates with giant magnetostrictive material layers *J. Appl. Phys.* **110**, 124514 (2011).

[107] L W Martin, S P Crane, Y-H Chu, M B Holcomb, M Gajek, M Huijben, C-H Yang, N Balke and R Ramesh. Multiferroics and magnetoelectrics: thin films and nanostructures. *J. Phys.: Condens. Matter.* 20 434220 (2008).

[108] B.B. Van Aken, A Meetsma, T.T.M Palstra. Hexagonal $YMnO_3$. *Acta Crystallogr*. C. 57 230 (2001).

[109] B.B. Van, T.T.M Palstra, A. Filippetti and N. Spaldin. The origin of ferroelectricity in magnetoelectric $YMnO_3$. Nature Materials. 3 **164** (2004).

[110] C. J. Fennie and K. M. Rabe. Ferroelectric transition in $YMnO_3$ from first principles. Phys. Rev. B . **72**, 100103R (2005).

[111] N. Fujimura, T. Ishida, T. Yoshimura, and T. Ito. Epitaxially grown $YMnO_3$ film: New candidate for nonvolatile memory devices. *Appl. Phys. Lett.* **69**, 1011 (1996).

[112] H.N. Lee, Y.T. Kim, Y.K. Park. Memory window of highly c-axis oriented ferroelectric $YMnO_3$ thin films. *Appl. Phys. Lett.* **74**, 3887 (1999).

[113] D Ch Yoo, J Y Leea, I S Kimb, Y T Kim. Microstructure control of $YMnO_3$ thin films on Si (100) substrates. Thin Solid Films **416** 62 (2002).

[114] D. Ito, N. Fujimura, T. Yoshimura, and T. Ito Ferroelectric properties of $YMnO_3$ epitaxial films for ferroelectric-gate field effect transistors. *J. Appl. Phys*. **93**, 5563 (2003).

[115] J. Dho, C.W. Leung, J.L. MacManus-Driscoll, M.G. Blamire. Epitaxial and oriented $YMnO_3$ film growth. by pulsed laser deposition. *J. Cryst. Growth* **267**, 548 (2004).

[116] K-T Kim, Ch-I Kim. The effects of drying temperature on the crystallization of $YMnO_3$ thin films prepared by sol-gel method using alkoxides. J. Eur. Ceram. Soc. 24 2613 (2004)

[117] K. Uusi-Esko, J. Malm, and M. Karppinen. Atomic Layer Deposition of Hexagonal and Orthorhombic $YMnO_3$ Thin Films. *Chem. Mater.*, **21**, 5691 (2009).

[118] I. Fina, L. Fàbrega, X. Martí, F. Sánchez, and J. Fontcuberta. Magnetic switch of polarization in epitaxial orthorhombic $YMnO_3$ thin films. *Appl. Phys. Lett*. **97**, 232905 (2010).

[119] H. Wadati, J. Okamoto, M. Garganourakis, V. Scagnoli, U. Staub, Y. Yamasaki, H. Nakao, Y. Murakami, M. Mochizuki, M. Nakamura, M. Kawasaki, and Y. Tokura. Origin of the Large Polarization in multiferroic $YMnO_3$ thin films revealed by Soft- and hard-X-Ray diffraction Phys. Rev. Lett. 108, 047203 (2012)

[120] C. Tabares-Muñoz, J.-P. Rivera, A. Bezinges, A. Monnier and H Schmid Measurement of the quadratic magnetoelectric effect on single crystalline $BiFeO_3$. *Jap. J. Appl. Phys.*, **24**, Suppl. 24-2, 1051 (1985).

[121] S. V. Kiselev, R. P. Ozerov, G. S. Zhdanov: Detection of magnetic order in ferroelectric $BiFeO_3$ by neutron diffraction, *Sov*. *Phys*. *Dokl*. **7**, 742 (1963).

[122] J.M. Moreau, C. Michel, R. Gerson and W.J. James, Ferroelectric $BiFeO_3$ x-ray and neutron diffraction study. *J. Phys. Chem*. Solids **32**, 1315 (1971).

[123] I. R. *Teague*, *R. Gerson*, and W. J. James. Dielectric hysteresis in single crystal $BiFeO_3$ Solid *State Commun*. **8**, 1073 (1970)

[124] G A Smolenskiĭ and I E Chupis. Ferroelectromagnets. *Sov. Phys. Usp.* 25 475 (1982).

[125] I. E. Dzyaloshinskii, Thermodynamic theory of weak ferromagnetism in antiferromagnetic substances. *Sov. Phys. JETP* **5**, 1259 (1957).

[126] V. R. Palkar, J. John, and R. Pinto Observation of saturated polarization and dielectric anomaly in magnetoelectric $BiFeO_3$ thin films. *Appl. Phys. Lett*. **80**, 1628 (2002).

[127] K. Y. Yun, M. Noda, and M. Okuyama. Prominent ferroelectricity of $BiFeO_3$ thin films prepared by pulsed-laser deposition. *Appl. Phys. Lett*. **83**, 3981 (2003).

[128] J. Li, J. Wang, M. Wuttig, R. Ramesh, N. Wang, B. Ruette, A. P. Pyatakov, A. K. Zvezdin, and D. Viehland. Dramatically enhanced polarization in (001), (101), and (111) $BiFeO_3$ thin films due to epitaxial-induced transitions. *Appl. Phys. Lett*. **84**, 5261 (2004).

[129] D. H. Wang, L. Yan, C. K. Ong, and Y. W. Du. $BiFeO_3$ film deposited on Si substrate buffered with $La_{0.7}Sr_{0.3}MnO_3$ electrode. *Appl. Phys. Lett*. **89**, 182905 (2006).

[130] V. V. Shvartsman, W. Kleemann, R. Haumont, and J. Kreisel Large bulk polarization and regular domain structure in ceramic BiFeO3. *Appl. Phys. Lett*. **90**, 172115 (2007).

[131] C. Ederer and N. A. Spaldin Influence of strain and oxygen vacancies on the magnetoelectric properties of multiferroic bismuth ferrite. *Phys. Rev. B*. **71**, 224103 (2005).

[132] S. K. Singh, H. Ishiwara, and K. Maruyama. Room temperature ferroelectric properties of Mn-substituted $BiFeO_3$ thin films deposited on Pt electrodes using chemical solution deposition *Appl. Phys. Lett*. **88**, 262908 (2006).

[133] B. Yu, M. Li, Z. Hu, L. Pei, D. Guo, X. Zhao, S. Dong. Enhanced multiferroic properties of the high-valence Pr doped $BiFeO_3$ thin film. *Appl. Phys. Lett*. **93**, 182909 (2008).

[134] D. Lee, Y-A Park, S M Yang, T K Song, Y Jo, N Hur, J H Jung and T W Noh. Suppressed magnetoelectric effect in epitaxially grown multiferroic $Pb(Zr_{0.57}Ti_{0.43})O_3$–$Pb(Fe_{2/3}W_{1/3})O_3$ solid-solution thin films. *J. Phys. D: Appl. Phys*. 43 455403 (2010)

[135] D. Lee, S.M Yang, Y. Jo, T.K Song. Room-temperature Multiferroic Properties of Pb(Zr$_{0.57}$Ti$_{0.43}$)O$_3$-Pb(Fe$_{0.67}$W$_{0.33}$)O$_3$ Solid-solution Epitaxial Thin Films. *J. Korean Phys. Soc.* 57, 1914 (2010)

[136] D. Sanchez, A. Kumar, N. Ortega, R. S. Katiyar, and J. F. Scott. Near-room temperature relaxor multiferroic. *Appl. Phys. Lett.* 97, 202910 (2010).

[137] R. E. Newnham, D. P. Skinner, and L. E. Cross. Connectivity and piezoelectric-pyroelectric composites. *Mater, Res. Bull.* 13, 525 (1978).

[138] S. X. Dong, J. R. Cheng, J. F. Li and D. Viehland. Enhanced magnetoelectric effects in laminate composites of Terfenol-D/Pb(Zr,Ti)O$_3$ under resonant drive. *Appl. Phys. Lett.* 83, 4812 (2003).

[139] A. D. Milliken, A. J. Bell, J. F. Scott, Dependence of breakdown field on dielectric (interelectrode) thickness in base-metal electroded multilayer capacitors. *Appl. Phys. Lett.* 90,112910 (2007).

[140] R. Martínez, A. Kumar, R. Palai, R. S. Katiyar, and J. F. Scott. Study of physical properties of integrated ferroelectric/ferromagnetic heterostructures. *Appl. Phys Lett..* 107, 114107 (2010).

[141] D. Palic, Z. Jaglicic, M. Jagodic, R. Blinc, J. Holc, M. Kosec, Z, Trontelj, Joint Eur. Mag. Symp. (JEMS), J. Phys. Conf. Series 303, 012065 (2011), ed. J. Spalek.

[142] H. Zheng, J. Wang, S. E. Lofland, Z. Ma, L. Mohaddes-Ardabili, T. Zhao, L. Salamanca-Riba, S. R. Shinde, S. B. Ogale, F. Bai, D. Viehland, Y. Jia, D. G. Schlom, M. Wuttig, A. Roytburd and R. Ramesh. Multiferroic BaTiO$_3$-CoFe$_2$O$_4$ Nanostructures. *Science* 303, 661-663 (2004).

[143] W. Peng; N. Lemee, M. Karkut et al., Appl. Phys. Lett. 94, 012509 (2009); W. Peng, N. Lemee, J. Holc et al., J. Mag. Mag. Mater. 321, 1754 (2009).

[144] W. Kleemann, V. V. Shvartsman, P. Borisov, and A. kania. Coexistence of Antiferromagnetic and Spin Cluster Glass Order in the Magnetoelectric Relaxor Multiferroic PbFe$_{0.5}$Nb$_{0.5}$O$_3$. *Phys. Rev. Lett.* 105, 257202 (2010).

[145] Ederer C. and Sapldin N.A. Recent progress in first-principles studies of magnetoelectric multiferroics. *Curr. Opin. Solid State Mater.* Sci. 9 128 (2005).

[146] C.W. Nan. G. Liu, Y. Lin. and H. Chen. Magnetic-Field-Induced Electric Polarization in Multiferroic Nanostructures. *Phys. Rev. Lett.* 94 197203 (2005).

[147] J.a Gao, X Zhu, W Liu, Z Zhang, J Cao, Ch Lin, D Zhu, and E. Liu. Ferroelectricity and ferromagnetism in (Pb,La)(Ca,Ti)O$_3$–La$_{0.67}$Sr$_{0.33}$MnO$_x$ multilayers. *Appl. Phys. Lett.* 78, 3869 (2001).

[148] K.-S. Chang, M. A. Aronova, C.-L. Lin, M. Murakami, M.-H. Yu, J. Hattrick-Simpers, O. O. Famodu, S. Y. Lee, R. Ramesh, M. Wuttig, I. Takeuchi, C. Gao, and L. A. Bendersky. Exploration of artificial multiferroic thin-film heterostructures using composition spreads. *Appl. Phys. Lett.* 84, 3091 (2004).

[149] P. Murugavel, D. Saurel, W. Prellier, Ch. Simon, and B. Raveau Tailoring of ferromagnetic Pr$_{0.85}$Ca$_{0.15}$MnO$_3$/ferroelectric Ba$_{0.6}$Sr$_{0.4}$TiO$_3$ superlattices for multiferroic properties Appl. Phys. Lett. 85, 4424 (2004).

[150] P. Murugavel, M. P. Singh, W. Prellier, B. Mercey, Ch. Simon, and B. Raveau. The role of ferroelectric-ferromagnetic layers on the properties of superlattice-based multiferroics. *J. Appl. Phys.* 97, 103914 (2005).

[151] M. A. Zurbuchen, T. Wu, S. Saha, J. Mitchell, and S. K. Streiffer. Multiferroic composite ferroelectric-ferromagnetic films *Appl. Phys. Lett.* 87, 232908 (2005)

[152] M. P. Singh, W. Prellier, Ch. Simon, and B. Raveau Magnetocapacitance effect in perovskite-superlattice based multiferroics. *Appl. Phys. Lett.* 87, 022505 (2005)

[153] Y. W. Li, J. L. Sun, J. Chen, X. J. Meng, and J. H. Chu. Structural, ferroelectric, dielectric, and magnetic properties of $BiFeO_3/Pb(Zr_{0.5},Ti_{0.5})O_3$ multilayer films derived by chemical solution deposition, *Appl. Phys. Lett.* **87**, 182902 (2005).

[154] J-P Zhou, H. He, Z Shi, and C-W Nan. Magnetoelectric $CoFe_2O_4/Pb(Zr_{0.52}Ti_{0.48})O_3$ double-layer thin film prepared by pulsed-laser deposition. *Appl. Phys. Lett*. **88**, 013111 (2006).

[155] M. P. Singh, W. Prellier, L. Mechin, Ch. Simon, and B. Raveau. Correlation between structure and properties in multiferroic $La_{0.7}Ca_{0.3}MnO_3/BaTiO_3$ superlattices. *J. Appl. Phys*. 99, 024105 (2006).

[156] R. V. Chopdekar and Y. Suzuki. Magnetoelectric coupling in epitaxial $CoFe_2O_4$ on $BaTiO_3$ *Appl. Phys. Lett*. **8**9, 182506 (2006).

[157] N. Ortega, P. Bhattacharya, R. S. Katiyar, P. Dutta, A. Manivannan, M. S. Seehra, I. Takeuchi, and S. B. Majumder. Multiferroic properties of $Pb(Zr,Ti)O_3/CoFe_2O_4$ composite thin films. *J. Appl. Phys*. 100, 126105 (2006).

[158] A. R Chaudhuri, R. Ranjith, S. B. Krupanidhi, R. V. K. Mangalam, and A. Sundaresan. Interface dominated biferroic $La_{0.6}Sr_{0.4}MnO_3/0.7Pb(Mg_{1/3}Nb_{2/3})O_3–0.3PbTiO_3$ epitaxial superlattices. *Appl. Phys. Lett*. **90**, 122902 (2007).

[159] W. Eerenstein, M. Wiora, J. L. Prieto, J. F. Scott, N. D. Mathur. Giant sharp and persistent converse magnetoelectric effects in multiferroic epitaxial heterostructures. *Nature Material*s **6**, 348 - 351 (2007).

[160] F Huang, X Lu, W Lin, W Cai, X Wu, Y Kan, H Sang, and J. Zhu Multiferroic properties and dielectric relaxation of $BiFeO_3/Bi_{3.25}La_{0.75}Ti_3O_{12}$ double-layered thin films. *Appl. Phys. Lett*. 90, 252903 (2007).

[161] S. Sahoo, S. Polisetty, Ch-Gg Duan, S S. Jaswal, E. Y. Tsymbal, and Ch. Binek. Ferroelectric control of magnetism in $BaTiO_3$/Fe heterostructures via interface strain coupling. *Phys. Rev. B* **76**, 092108 (2007).

[162] Y. G. Ma, W. N. Cheng, M. Ning, and C. K. Ong. Magnetoelectric effect in epitaxial $Pb(Zr_{0.52}Ti_{0.48})O_3/La_{0.7}Sr_{0.3}MnO_3$ composite thin film. *Appl. Phys. Lett*. **90**, 152911 (2007).

[163] N. Ortega, Ashok Kumar, R. S. Katiyar, and J. F. Scott. Maxwell-Wagner space charge effects on the $Pb(Zr,Ti)O_3–CoFe_2O_4$ multilayers . *Appl. Phys. Lett*. **91**, 102902 (2007).

[164] J. X. Zhang, J. Y. Dai, C. K. Chow, C. L. Sun, V. C. Lo, and H. L. W. Chan Magnetoelectric coupling in $CoFe_2O_4/SrRuO_3/Pb(Zr_{0.52}Ti_{0.48})O_3$ heteroepitaxial thin film structure. *Appl. Phys. Lett*. **92**, 022901 (2008).

[165] H-C He, J. Ma, J Wang, and C-W Nan. Orientation-dependent multiferroic properties in $Pb(Zr_{0.52}Ti_{0.48})O_3–CoFe_2O_4$ nanocomposite thin films derived by a sol-gel processing. *J. Appl. Phys*. **103**, 034103 (2008).

[166] R Lin, J Liao, L Hung, and T Wu. Effect of the $CoFe_2O_4$ thin film thickness on multiferroic property of (00*l*)-oriented $Pb(Zr_{0.5}Ti_{0.5})O_3/CoFe_2O_4/Pb(Zr_{0.5}Ti_{0.5})O_3$ trilayer structure. *J. Appl. Phys*. **103**, 07E320 (2008).

[167] M. Liu, O. Obi, J. Lou, S. Stoute, J. Y. Huang, Z. Cai[3], K. S. Ziemer, and N. X. Sun. Spin-spray deposited multiferroic composite $Ni_{0.23}Fe_{2.77}O_4/Pb(Zr,Ti)O_3$ with strong interface adhesion *Appl. Phys. Lett*. **92**, 152504 (2008).

[168] C. H. Sim, A. Z. Z. Pan, and J. Wang. Thickness and coupling effects in bilayered multiferroic $CoFe_2O_4/Pb(Zr_{0.52}Ti_{0.48})O_3$ thin films *J. Appl. Phys*. **103**, 124109 (2008).

[169] R. Y. Zheng, J. Wang, and S. Ramakrishna. Electrical and magnetic properties of multiferroic $BiFeO_3/CoFe_2O_4$ heterostructure. *J. Appl. Phys*. **104**, 034106 (2008).

[170] N. Ortega, Ashok Kumar, P. Bhattacharya, S. B. Majumder, and R. S. Katiyar. Impedance spectroscopy of multiferroic PbZr$_x$Ti$_{1-x}$O$_3$/CoFe$_2$O$_4$ layered thin films. *Phys. Rev. B* **77**, 014111 (2008).

[171] Z. Li, Y. Wang, Y. Lin, and C-W Nan. Evidence for stress-mediated magnetoelectric coupling in multiferroic bilayer films from magnetic-field-dependent Raman scattering. *Phys. Rev. B*. **79**, 180406(R) (2009).

[172] A. Kumar, R. S. Katiyar, R N Premnath, C. Rinaldi and J. F. Scott. Strain-induced artificial multiferroicity in Pb(Zr$_{0.53}$Ti$_{0.47}$)O$_3$/Pb(Fe$_{0.66}$W$_{0.33}$)O$_3$ layered nanostructure at ambient temperature. *J. Mater. Sci.*, 44, 5113-5119 (2009).

[173] N. Ortega, Ashok Kumar, C. Rinaldi and Ram S. Katiyar. Dynamic magneto-electric multiferroics PZT/CFO multilayered nanostructure. *J Mater Sci* 44:5127–5142 (2009).

[174] P. Padhan, P. LeClair, A. Gupta, M. A. Subramanian and G. Srinivasan. Magnetodielectric effect in Bi$_2$NiMnO$_6$–La$_2$NiMnO$_6$ superlattices. *J. Phys.: Condens. Matter* **21** 306004 (2009).

[175] S. Dussan, A. Kumar, J. F. Scott, and R.S. Katiyar. Magnetic effects on dielectric and polarization behavior of multiferroic heterostructures. *Appl. Phys. Lett.* **96**, 072904 (2010);

[176] S. Dussan, A. Kumar, J. F. Scott, S. Priya, and R. S. Katiyar. Room temperature multiferroic effects in superlattice nanocapacitors. *Appl. Phys. Lett.* **97**, 252902 (2010).

[177] S. Dussan, A Kumar, R S Katiyar, S. Priya and J. F. Scott. Magnetic control of ferroelectric interfaces *J. Phys.: Condens. Matter* **23** 202203 (2011).

[178] R Martínez, A Kumar, R Palai, J F Scott and R S Katiyar. Impedance spectroscopy analysis of Ba$_{0.7}$Sr$_{03}$TiO$_3$/La$_{0.7}$Sr$_{0.3}$MnO$_3$ heterostructure. *J. Phys. D: Appl. Phys*. **44** 105302 (2011).

[179] R Martínez, A Kumar, R Palai, R S Katiyar and G. Srinivasan. Observation of strong magnetoelectric effects in Ba$_{0.7}$Sr$_{0.3}$TiO$_3$/La$_{0.7}$Sr$_{0.3}$MnO$_3$ thin film heterostructures. *J. Appl. Phys*. (accepted).

[180] M. Plekh and M. Tyunina. Ferroelectric domains in epitaxial PbZr$_{0.65}$Ti$_{0.35}$O$_3$/La$_{0.5}$Sr$_{0.5}$CoO$_3$ heterostructures. *Appl. Phys. Lett*. **97**, 062902 (2010).

[181] C. A. F. Vaz, J. Hoffman, Y. Segal, M. S. J. Marshall, J. W. Reiner, Z. Zhang, R. D. Grober, F. J. Walker, and C. H. Ahn. Control of magnetism in Pb(Zr$_{0.2}$Ti$_{0.8}$)O$_3$/La$_{0.8}$Sr$_{0.2}$MnO$_3$ multiferroic heterostructures . *Appl. Phys*. **109**, 07D905 (2011).

[182] Z. Li, J Hu, L. Shu, Y. Zhang, Y. Gao, Y. Shen, Y. Lin, and C. W. Nan A simple method for direct observation of the converse magnetoelectric effect in magnetic/ferroelectric composite thin films. *J. Appl. Phys*. **110**, 096106 (2011).

[183] T. Yu, H. Naganuma, W. X. Wang, Y. Ando, and X. F. Han. Annealing temperature dependence of exchange bias in BiFeO$_3$/CoFe bilayers. *J. Appl. Phys*. **111**, 07D908 (2012).

[184] Ch-J Hsu, J. L. Hockel, and G P Carman. Magnetoelectric manipulation of domain wall configuration in thin film Ni/[Pb(Mn$_{1/3}$Nb$_{2/3}$)O$_3$]$_{0.68}$-[PbTiO$_3$]$_{0.32}$ (001) heterostructure. *Appl. Phys. Lett.* **100**, 092902 (2012).

[185] G. Venkataiah, Y. Shirahata, I. Suzuki, M. Itoh, and T. Taniyama Strain-induced reversible and irreversible magnetization switching in Fe/BaTiO$_3$ heterostructures. *J. Appl. Phys*. **111**, 033921 (2012).

[186] Z Li, J Hu, L Shu, Y Gao, Y Shen, Y Lin, and C. W. Nan. Thickness-dependent converse magnetoelectric coupling in bi-layered Ni/PZT thin films. *J. Appl. Phys*. **111**, 033918 (2012).

[187] Y. Yang, M. M Yang, Z. L. Luo, H Huang, H Wang, J. Bo, ChHu, G Pan, Y Yao, Y Liu, X. G. Li, Sen Zhang, Y. G. Zhao, and C. Gao. Large anisotropic remnant magnetization tunability in (011)-La$_{2/3}$Sr$_{1/3}$MnO$_3$/0.7Pb(Mg$_{2/3}$Nb$_{1/3}$)O$_3$-0.3PbTiO$_3$ multiferroic epitaxial heterostructures *Appl. Phys. Lett*. **100**, 043506 (2012).

[188] H. Zheng, J. Wang, L. Mohaddes-Ardabili, M. Wuttig, L. Salamanca-Riba, D. G. Schlom, and R. Ramesh. Three-dimensional heteroepitaxy in self-assembled $BaTiO_3$–$CoFe_2O_4$ nanostructures. *Appl. Phys. Lett*. **85**, 2035 (2004).

[189] H. Zheng, J Kreisel, Y-H Chu, R. Ramesh, and L. Salamanca-Riba. Heteroepitaxially enhanced magnetic anisotropy in $BaTiO_3$–$CoFe_2O_4$ nanostructures. *Appl. Phys. Lett*. **90**, 113113 (2007).

[190] J. Li, I. Levin, J. Slutsker, V. Provenzano, P. K. Schenck, R. Ramesh, J. Ouyang, and A. L. Roytburd. Self-assembled multiferroic nanostructures in the $CoFe_2O_4$-$PbTiO_3$ system. *Appl. Phys. Lett.* **87**, 072909 (2005).

[191] R. Muralidharan, N. Dix, V. Skumryev, M. Varela, F. Sánchez, and J. Fontcuberta. Synthesis, structure, and magnetic studies on self-assembled $BiFeO_3$–$CoFe_2O_4$ nanocomposite thin films. *J. Appl. Phys.* **103**, 07E301 (2008).

[192] N. M. Aimon, D Kim, H- K Choi, and C. A. Ross. Deposition of epitaxial $BiFeO_3$/$CoFe_2O_4$ nanocomposites on (001) $SrTiO_3$ by combinatorial pulsed laser deposition. *Appl. Phys. Lett*. **100**, 092901 (2012).

[193] R. Hajndl, J. Sanders, H. Srikanth, and N. J. Dudney. Growth and characterization of BSTO/hexaferrite composite thin films. *Appl. Phys Lett*. **93**, 7999 (2003).

[194] J. G. Wan, X. W. Wang, Y. J. Wu, M. Zeng, Y. Wang, H. Jiang, W. Q. Zhou, G. H. Wang, and J.-M. Liu. Magnetoelectric $CoFe_2O_4$–$Pb(Zr,Ti)O_3$ composite thin films derived by a sol-gel process. *Appl. Phys. Lett*. **86**, 122501 (2005).

[195] J. Barbosa, B. Almeida, J. A. Mendes, A. G. Rolo, . P. Araújo. X-ray diffraction and Raman study of nanogranular $BaTiO_3$–$CoFe_2O_4$ thin films deposited by laser ablation on Si/Pt substrates. *Phys. Stat. sol. (a)* **204**, 1731–1737, (2007).

[196] X. L. Zhong, J. B. Wang, M. Liao, G. J. Huang, S. H. Xie, Y. C. Zhou, Y. Qiao, and J. P. He. Multiferroic nanoparticulate $Bi_{3.15}Nd_{0.85}Ti_3O_{12}$–$CoFe_2O_4$ composite thin films prepared by a chemical solution deposition technique. *Appl. Phys. Lett*. **90**, 152903 (2007).

[197] J H. Park, H M. Jang, H S. Kim, Ch G. Park, and S G. Lee. Strain-mediated magnetoelectric coupling in $BaTiO_3$-Co nanocomposite thin films *Appl. Phys. Lett*. **92**, 062908 (2008).

[198] J X Zhang, J Y Dai, W Lu1, H L W Chan, B Wu and D X Li. A novel nanostructure and multiferroic properties in $Pb(Zr_{0.52}Ti_{0.48})O_3$/$CoFe_2O_4$ nanocomposite films grown by pulsed-laser deposition. J. Phys. D: Appl. Phys. 41 235405 (2008).

[199] N. Mathur. Materials science: A desirable wind up. Nature. 454, 591 (2008).

[200] M. Bibes and A. Barthélémy. Multiferroics: Towards a magnetoelectric memory. *Nature Materials*. **7**, 425 (2008).

[201] N. A. Spaldin and M. Fiebig. The Renaissance of Magnetoelectric Multiferroics. S*cience* 309, 391-392 (*2005*).

[202] T. Tybell, C. H. Ahn, J.-M. Triscone. *Appl. Phys. Lett.* **75**, 856 (1999)

[203] S. Yuasa, T. Nagahama, A. Fukushima, Y. Suzuki, and K. Ando. *Nat. Mat.* **3**: 868 (2004).

[204] E. Y. Tsymbal and H. Kohlstedt. Tunneling Across a Ferroelectric. *Science*, 313, 181 (2006).

[205] M. Gajek, M. Bibes, S. Fusil, K. Bouzehouane, J. Fontcuberta, A. Barthélémy and A. Fert. Tunnel junctions with multiferroic barriers. *Nature Materials.* **6,** 296 (2007).

[206] M. Gajek, M. Bibes, A. Barthélémy, K. Bouzehouane, S. Fusil, M. Varela, J. Fontcuberta, and A. Fert. *Phys. Rev. B***. 72**, 020406(R) (2005).